\documentclass[submission, Phys]{SciPost}

\usepackage{amssymb}
\usepackage{graphicx}
\usepackage{xcolor}
\usepackage{dcolumn}
\usepackage{bm}
\usepackage{dsfont}
\usepackage{appendix}
\usepackage{placeins}

\usepackage{booktabs}

\newcommand{\HAF}{H_{\mathrm{AF}}}
\newcommand{\HK}{H_{K}}
\newcommand{\UF}{U_F}
\newcommand{\aE}{\bar{E}}

\begin{document}

\begin{center}{\Large \textbf{\color{scipostdeepblue}{
Stroboscopic stability of a Floquet chiral spin liquid beyond the folding
frequency
}}}\end{center}

\begin{center}{\large
Didier Poilblanc\textsuperscript{1$\star$}
}\end{center}

\begin{center}
{\bf 1} Universit\'e de Toulouse, CNRS, Laboratoire de Physique Th\'eorique,
Toulouse, France
\\[\baselineskip]
$\star$ \href{mailto:didier.poilblanc@gmail.com}{\small didier.poilblanc@gmail.com}
\end{center}

\section*{\color{scipostdeepblue}Abstract}
{\bf
We study the two-step Floquet dynamics of the chiral $J_1$--$J_2$--$K$ Heisenberg
model on the $4\times4$ torus, alternating its non-chiral Heisenberg part $\HAF$
and its chiral plaquette part $\HK$, at the parameter point where the static model
(recovered in the infinite frequency limit) hosts a quasi-degenerate, spectrally isolated chiral-spin-liquid (CSL) topological
doublet. The one-period propagator is computed exactly in all momentum/rotation
symmetry sectors. It is shown that, decreasing the frequency, the topological doublet survives the drive far beyond the
frequency $\omega_{\mathrm{res}}\simeq 11.5 J_1$ at which folded states first cross it in
quasienergy: the time-averaged energy of the Floquet eigenstates, which orders the
folded Floquet spectrum, shows that the doublet remains the isolated bottom of the spectrum
down to $\omega\simeq6J_1$, while stroboscopic time evolution over thousands of
periods shows no heating for $\omega\gtrsim \omega_{\mathrm{res}}$ and only slow absorption below.
The quasienergy resonances that occur in the folded regime are invisible in the
average energy, identifying them as parametrically weak avoided crossings. We argue
that this mechanism --- stability controlled by local energy scales rather than by
the extensive many-body bandwidth --- is of the type expected to survive in the
thermodynamic limit, where a prethermal Floquet CSL should persist for
$\omega$ above a threshold set by local scales, with heating times exponentially
long in $\omega/J_1$ --- a conjecture that larger clusters can now test.
Consistently, the optimal $D=3$ chiral
PEPS of the static problem still describes the driven doublet deep in the folded
regime, with an essentially unchanged local tensor.
Finally, we show that the drive is realizable on existing quantum simulators, by
giving an explicit digital pulse sequence built solely from XY-type two-spin
gates on the four bond colours of the square lattice, which realizes the two
half-steps independently with an infidelity of $1.5\times10^{-5}$ per period.
}

\vspace{\baselineskip}

\noindent\textcolor{scipostdeepblue}{\rule{\textwidth}{1pt}}
\tableofcontents
\noindent\textcolor{scipostdeepblue}{\rule{\textwidth}{1pt}}
\vspace{\baselineskip}

\section{Introduction}
\label{sec:intro}

Chiral spin liquids (CSL) --- the spin analogues of bosonic fractional quantum Hall
states first envisioned by Kalmeyer and Laughlin \cite{kalmeyer1987,wen1989} --- are
among the most striking phases of frustrated quantum magnetism: gapped bulk,
topological ground-state degeneracy on the torus, fractional (semionic) excitations
and chiral edge modes. In static $SU(2)$-invariant microscopic models they are
typically stabilized by explicitly breaking time reversal with imaginary cyclic
plaquette terms. The canonical such model --- nearest- and next-nearest-neighbour
Heisenberg couplings supplemented by a scalar-chirality term --- descends from the
local truncation of the Kalmeyer--Laughlin parent Hamiltonian constructed by
Nielsen, Sierra and Cirac \cite{nielsen2013}, and its phase diagram has been
mapped with PEPS \cite{poilblanc2017}. In all such constructions, however, the
time-reversal-breaking interaction has to be put in by hand, and the chiral
phase survives only in a restricted window of couplings.

Periodic driving offers an alternative and, in several respects, richer route.
Floquet engineering can generate effective chiral interactions dynamically, as
shown for the driven square-lattice Heisenberg antiferromagnet
\cite{mambrini2024,pmg2024}, and can
A closely related and earlier route is optical: in frustrated Mott insulators a
circularly polarized field breaks time reversal dynamically and generates a
scalar-chirality term at order $1/\omega$, which was shown to stabilize a
photo-induced CSL \cite{claassen2017} and, in parallel work, to allow the spin
chirality to be probed and controlled directly \cite{kitamura2017}. The drive
studied here differs in that the chiral term is present microscopically rather
than generated perturbatively, but the underlying question --- whether such a
dynamically stabilized chiral phase survives the heating --- is the same.
Periodic driving can moreover
produce \emph{anomalous} chiral phases with no static counterpart --- a
possibility first identified in periodically driven two-dimensional band systems,
where protected chiral edge states exist even when all Chern numbers vanish
\cite{rudner2013}, and extended to interacting many-body systems with the chiral
(``swap'') Floquet phases of driven bosons \cite{po2016} --- as shown recently for
a driven ``Swap'' spin model hosting an anomalous Floquet CSL \cite{detuning}.
The price is the threat of heating: a generic interacting system absorbs energy from
the drive and flows towards an infinite-temperature state. High-frequency expansions
and rigorous prethermalization bounds \cite{mori2016,kuwahara2016,abanin2017}
guarantee exponentially slow heating when $\hbar\omega$ exceeds the \emph{local}
energy scales, and underpin the notion of a prethermal \emph{phase} with its own
symmetry structure \cite{else2017}, but a naive spectral criterion is far more restrictive: as soon as
$\hbar\omega$ is smaller than the (extensive) many-body bandwidth, the Floquet zone
folds and the topological doublet is crossed by a macroscopic number of high-energy
states. Since folding is unavoidable in the thermodynamic limit at any fixed
frequency, the practical existence of a Floquet CSL hinges on a sharp question:
\emph{is spectral folding actually harmful, or is stability controlled by local
scales only?}

Answering this question requires an observable that remains meaningful once the
Floquet zone has folded. Such a tool was provided recently by the geometric Floquet
theory of Schindler and Bukov \cite{schindler2025}, which reformulates Floquet
dynamics in terms of parallel-transported eigenstates and, in doing so, promotes
the \emph{time-averaged energy} of each Floquet eigenstate to the natural,
folding-free generalization of the energy. This concept, already instrumental in
the analysis of the driven Swap model of Ref.~\cite{detuning}, provides the key
diagnostic of the present work: it orders the folded Floquet spectrum at any
frequency and thereby turns the stability question into a quantitative one.

Here we address this question quantitatively in the simplest setting: a two-step
(split-step) drive that alternates the non-chiral Heisenberg part and the chiral
plaquette part of the optimized $J_1$--$J_2$--$K$ Hamiltonian, at the parameter point
where the static model hosts its best CSL doublet. The one-period propagator is
computed \emph{exactly} on the $4\times4$ torus, in all symmetry sectors, for
arbitrary period $T$ (Sec.~\ref{sec:model}). We follow the topological doublet with
four independent diagnostics: quasienergy splitting and isolation on the Floquet
circle (Sec.~\ref{sec:quasi}); the average energy of the Floquet eigenstates
\cite{schindler2025,detuning} discussed above (Sec.~\ref{sec:ae}); the variational $D=3$ chiral PEPS description of the driven
doublet (Sec.~\ref{sec:peps}); and stroboscopic time evolution over thousands of
periods (Sec.~\ref{sec:strobe}). All of them consistently show that the doublet survives far
beyond the first folding resonances, and that its eventual destruction is governed
by a local, size-independent frequency scale --- the basis of our argument
(Sec.~\ref{sec:discussion}) that a prethermal Floquet CSL survives in the
thermodynamic limit.

\section{Model and drive}
\label{sec:model}

We consider the static chiral $J_1$--$J_2$--$K$ Heisenberg model on the square lattice, which is split as
$H=\HAF+\HK$ with
\begin{align}
\HAF &= J_1 \sum_{\langle ij\rangle} \mathbf{S}_i\cdot\mathbf{S}_j
\;+\; J_2 \sum_{\langle\langle ij\rangle\rangle} \mathbf{S}_i\cdot\mathbf{S}_j ,
\label{eq:HAF}\\
\HK &= K \sum_{p} \left[ \cos\theta \left( P_p + P_p^{-1} \right)
+ i \sin\theta \left( P_p - P_p^{-1} \right) \right],
\label{eq:HK}
\end{align}
where $P_p$ cyclically permutes the four spins of plaquette $p$ and $J_1=1$ sets the
unit of energy. 
The model used here extends the canonical family introduced in
Ref.~\cite{nielsen2013}: besides the imaginary chiral term it retains the
\emph{real} part $\cos\theta\,(P_p+P_p^{-1})$ of the cyclic permutation --- a
four-site ring exchange. The reason is one of tunability. Even within the region
hosting a well-formed CSL, a topological doublet that is simultaneously
quasi-degenerate \emph{and} spectrally isolated is found, on finite systems, only
in carefully tuned corners of parameter space, and the extra term widens the
window within which such a point can be located.

The cluster used throughout the main text is the $4\times4$ torus ($N=16$ sites,
periodic boundary conditions). There, a well-formed doublet requires fine-tuned
values of both the frustrating coupling $J_2$ and the angle $\theta$, at which
the real and chiral parts of the cyclic term carry comparable weight
(Appendix~\ref{app:static}). The real exchange is thus part of what makes the
sweet spot available at this size; it is not in fact necessary on larger
clusters (Appendix~\ref{app:larger}).

Throughout we work at the ``sweet spot''
\begin{equation}
(\theta/\pi,\,K,\,J_2) = (0.325,\;0.20,\;0.40),
\label{eq:sweetspot}
\end{equation}
determined in Appendix~\ref{app:static}, at which the static model possesses
the two quasi-degenerate singlet ground states expected for a CSL on a torus ---
the lowest $S=0$ states at
$\mathbf{k}=(0,0)$ with $C_4$ eigenvalue $+1$ ($A$) and $-1$ ($B$) --- split by
$\Delta_{AB}=0.025$ and isolated from all other states by
$\Delta_{\mathrm{iso}}=0.315$. These quantum numbers are specific to this
cluster: on the tori of $N=20$ and $26$ sites considered in
Appendix~\ref{app:larger} the doublet is labelled differently.

The system is driven by alternating the two parts of $H$ over a period $T$,
\begin{equation}
\boxed{\;\UF(T) \;=\; e^{-i\frac{T}{2}\HK}\; e^{-i\frac{T}{2}\HAF}\;},
\label{eq:UF}
\end{equation}
i.e.\ half a period of Heisenberg evolution followed by half a period of purely chiral
evolution. The stroboscopic dynamics generated by (\ref{eq:UF}) can always be viewed
as the evolution under an effective \emph{static} Hamiltonian, the Floquet
Hamiltonian $H_F(T)$, defined by
$
\UF(T) \;=\; e^{-i\,T\,H_F(T)} .
$
Note that, $\UF$ being unitary, $H_F$ is defined only up to the
branch choice of the matrix logarithm --- the origin of the quasienergy
``folding'' central to this work (Sec.~\ref{sec:quasi}). In the high-frequency
limit $H_F$ reduces to the time average,
$H_F \to \overline{H} = \tfrac12(\HAF+\HK) = H/2$: all static energies and
gaps appear halved, and the drive connects smoothly to the static sweet-spot physics.

Because the interest of such a protocol ultimately rests on whether it can be run,
two appendices address implementation. Appendix~\ref{app:larger} extends the static
exact diagonalization to $C_4$-symmetric tori of $N=20$ and $26$ sites and shows
that the two ingredients which are hardest to engineer --- the real part
$\cos\theta\,(P_p+P_p^{-1})$ of the cyclic term, and the frustrating second-neighbour
coupling $J_2$ --- may both be dispensed with: at $N=26$ the purely chiral point
$\theta=\pi/2$ hosts a doublet \emph{better} isolated than the fine-tuned sweet spot
used here. Appendix~\ref{app:pulses} then gives an explicit digital realization of
$\UF(T)$ itself, assembled entirely from XY-type two-spin gates of the kind
natively available on Rydberg-atom, polar-molecule and NV platforms
\cite{nishad2023}. The two are
linked: switching off the real part of the cyclic term is precisely what brings the
chiral half-step within reach of a two-body gate set.

It is also useful to contrast the drive (\ref{eq:UF}) with the Floquet-engineering
protocols studied previously for the same square-lattice antiferromagnet, with a
sinusoidal \cite{mambrini2024,pmg2024} 
waveform. In those schemes
the chiral plaquette interaction is absent from $H(t)$ at \emph{all} times: the
Hamiltonian is purely Heisenberg, with the four nearest-neighbor bond families
modulated with relative dephasings of $\pi/2$, and the chiral term
$i(P_p-P_p^{-1})$ only emerges at order $1/\omega$ of the high-frequency
expansion, with an effective coupling $J_F=J^2/4\omega$ generated by the
commutator of the bond modulations. The CSL is then a genuinely \emph{dynamical}
object: it disappears in the infinite-frequency limit and requires a compromise
between a large $\omega$ (to avoid heating) and a small $\omega$ (to achieve a
sizable $J_F$), which confines it to a finite window of frequencies
\cite{pmg2024}. The two-step drive (\ref{eq:UF}) is different in both respects:
the chiral term is explicitly present during half of each period, so the CSL
doublet already exists in the $\omega\to\infty$ limit --- where $H_F\to H/2$ is
the optimized static Hamiltonian --- and no $1/\omega$ trade-off is involved;
the issue is instead the \emph{stability} of this doublet as the frequency is
lowered into the folded regime. The piecewise-constant protocol has the further
technical advantage that $\UF$ factorizes exactly into two sector-diagonalized
exponentials, giving access to arbitrary periods without Trotter error and to
an exact evaluation of the average energy (Sec.~\ref{sec:ae}).

Both $\HAF$ and $\HK$ commute with lattice translations, $C_4$ rotations and total
spin, so $\UF$ is block diagonal in the nine symmetry sectors of the $4\times4$ torus
($\mathbf{k}=(0,0)$ resolved into $C_4=\pm1,\pm i$; and the momentum families of
$(\pi,\pi)$, $(\pi,0)$, $(\pi/2,0)$, $(\pi,\pi/2)$, $(\pi/2,\pi/2)$, of respective
multiplicities $1,2,4,4,4$; the sector dimensions, weighted by the multiplicity of each
momentum star, sum to $\binom{16}{8}=12870$, the dimension of the $S^z=0$
subspace). Within each sector $\HAF$ and $\HK$ are diagonalized once and
$\UF(T)$ is assembled \emph{exactly} for any $T$; its full eigendecomposition
follows from a complex Schur factorization. The implementation was
verified against an independent Pad\'e construction of the two exponentials
($10^{-15}$), unitarity and $[\UF,S^2]=0$ hold to machine precision, the $T\to0$
spectrum reproduces that of $\overline{H}$, and all static benchmarks of
Appendix~\ref{app:static} are reproduced exactly.

\section{Quasienergy spectrum and the folding scales}
\label{sec:quasi}

The natural basis for the stroboscopic dynamics is the \emph{Floquet eigenbasis},
i.e.\ the eigenstates of the one-period propagator,
\begin{equation}
\UF(T)\,|u_n\rangle \;=\; e^{-i\varepsilon_n T}\,|u_n\rangle ,
\qquad
\varepsilon_n \in \left(-\pi/T,\;\pi/T\right] ,
\label{eq:floquetbasis}
\end{equation}
which are also the eigenstates of $H_F$.
Since the eigenvalues
of the unitary $\UF$ are pure phases, they determine the \emph{quasienergies}
$\varepsilon_n$ only modulo $\omega=2\pi/T$ --- hence the prefix ``quasi'': under
a time-\emph{periodic} Hamiltonian, energy is conserved only up to an integer
number of drive quanta $\hbar\omega$, in the same way as the quasimomentum of
Bloch theory is conserved only modulo reciprocal-lattice vectors under a
spatially periodic potential. All quasienergies can therefore be brought
(``folded'') into a single Floquet zone, chosen in Eq.~(\ref{eq:floquetbasis})
as $(-\pi/T,\pi/T]$, or equivalently pictured as points on a circle of
circumference $2\pi/T$.

\begin{figure}[t]
\centering
\includegraphics[width=0.95\textwidth]{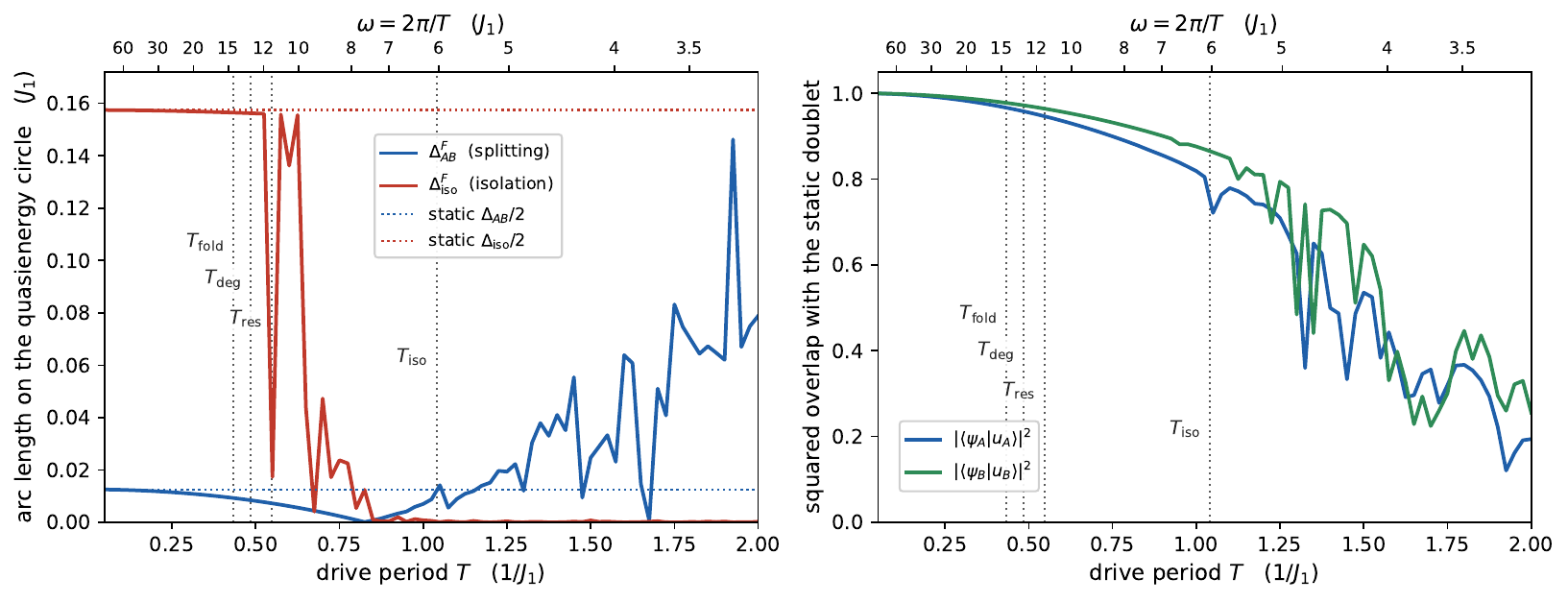}
\caption{Left: doublet splitting $\Delta^F_{AB}$ and isolation
$\Delta^F_{\mathrm{iso}}$ measured on the quasienergy circle, versus the drive
period $T$ (upper axis: $\omega=2\pi/T$). Horizontal dotted lines: static values
rescaled by $1/2$ (high-frequency limit $H_F\to H/2$); vertical dotted lines: the
key periods of Eqs.~(\ref{eq:wfold})--(\ref{eq:wres}) and Sec.~\ref{sec:ae}.
At small $T$ the isolation sits on a plateau at $0.1571$, indistinguishable from
the static value $\Delta_{\mathrm{iso}}/2=0.1574$, and the level that sets it is
the same $\mathbf{k}=(\pi,\pi)$ singlet that is the nearest competitor of the
static model --- we have checked that this intruder supplies the minimum
throughout the plateau, $T\le0.40$. Right: squared overlaps of the Floquet
doublet eigenstates with the static CSL doublet.}
\label{fig:doublet}
\end{figure}

Figure~\ref{fig:doublet} tracks the topological doublet in this quasienergy
spectrum as a function of the period (the full spectrum of all $12870$ levels,
from which the doublet is extracted, is shown in Appendix~\ref{app:fullspec},
Fig.~\ref{fig:fullspec}). The doublet members are identified in their
symmetry sectors by maximal overlap with the static states $\psi_{A,B}$ (see right panel).
Because the folded quasienergies live on a circle of circumference $2\pi/T$, with no
global ``above/below'' ordering, the splitting and isolation of
Appendix~\ref{app:static} must be recast as arc lengths on that circle. Writing
$d(x,y)=\min_{p\in\mathbb{Z}}|x-y-p\,\tfrac{2\pi}{T}|$ for the circle distance, the
splitting $\Delta^F_{AB}=d(\varepsilon_A,\varepsilon_B)$ is the shorter arc between the
two doublet quasienergies, while the isolation
\begin{equation}
\Delta^F_{\mathrm{iso}}
=\min_{n\notin\{A,B\}}\,
\min\!\big(d(\varepsilon_n,\varepsilon_A),\,d(\varepsilon_n,\varepsilon_B)\big)
\label{eq:iso}
\end{equation}
is the shortest arc separating \emph{either} doublet member from the nearest of all
other $12870$ levels (every symmetry sector and spin), each intruder being measured to
whichever of $\varepsilon_A,\varepsilon_B$ is the closer. This unsigned distance
coincides with the static gap when $\varepsilon_A,\varepsilon_B$ sit together at the
bottom of the unfolded spectrum (high frequency), but --- unlike the signed static
definition --- stays well defined once the spectrum folds and intruders wind around the
circle, and it is by construction insensitive to how far apart $A$ and $B$ themselves
lie (that separation being the splitting $\Delta^F_{AB}$).
Two remarkable features emerge at high frequency. First, the drive \emph{reduces} the
doublet splitting below its static value: $\Delta^F_{AB}$ decreases monotonically from
$\Delta_{AB}/2=0.0126$ at $T\to0$ to $0.0077$ at $T=0.525$, while the isolation stays
pinned at its static value $\Delta_{\mathrm{iso}}/2=0.157$ with the same
$\mathbf{k}=(\pi,\pi)$ singlet intruder as in the static model. Second, the overlaps
with the static doublet remain large, $|\langle\psi_A|u_A\rangle|^2=0.95$ and
$|\langle\psi_B|u_B\rangle|^2=0.97$ at $T=0.5$.

The unfolded quasienergies span
$[\varepsilon_{\min},\varepsilon_{\max}]=[-4.196,\,+7.272]$, of spread $W=11.47$.
We stress that these are the eigenvalues of $\overline{H}=H/2$, \emph{not} of the
static $H$: with the convention (\ref{eq:UF}) the quasienergies tend to the
spectrum of $\overline{H}$ as $T\to0$, so every static energy and gap is compared
after rescaling by $1/2$, and $W$ below always denotes the spread of
$\overline{H}$. Where the spread of the static $H$ itself is needed --- it is the
natural normalization for the absorbed energy of Sec.~\ref{sec:strobe}, which is
measured with $H$ --- we write $W_H=2W=22.94$.
Two analytic scales organize the fate of the doublet, and both are confirmed
sharply by the numerics:
\begin{align}
\omega_{\mathrm{fold}} &= 2\,\varepsilon_{\max} = 14.5\,J_1
\qquad (T_{\mathrm{fold}}=\pi/\varepsilon_{\max}=0.432),
\label{eq:wfold}\\
\omega_{\mathrm{res}} &= \varepsilon_{\max}-\varepsilon_{\min} = W = 11.5\,J_1
\qquad (T_{\mathrm{res}}=2\pi/W=0.548).
\label{eq:wres}
\end{align}
At $\omega_{\mathrm{fold}}$ the top of the many-body spectrum exits the Floquet zone
(folding onset); at $\omega_{\mathrm{res}}$ the first folded state reaches the
quasienergy of the doublet, which sits at the very bottom of the spectrum --- the
observed first resonance at $T=0.55$ is indeed produced by a high-energy $A$-sector
singlet folding down. In the window
$\omega_{\mathrm{res}}<\omega<\omega_{\mathrm{fold}}$ the spectrum is already folded
but the doublet is untouched: its isolation keeps its static value, its splitting
keeps shrinking, and (Sec.~\ref{sec:strobe}) it does not heat. Below
$\omega_{\mathrm{res}}$, individual folded states sweep through the doublet as a comb
of \emph{narrow} resonances, between which the quasienergy isolation recovers its full
value --- a first hint that these crossings are only weakly avoided.

\section{Average energy: ordering the folded spectrum}
\label{sec:ae}

Quasienergies are only defined modulo $2\pi/T$, so once the spectrum folds they
provide no notion of ``low-lying'' states. Following Refs.~\cite{schindler2025,detuning}
--- the average energy was introduced in the geometric Floquet theory of
Ref.~\cite{schindler2025}, and used for a driven spin model in Ref.~\cite{detuning}
[Eq.~(C1) and Fig.~5 therein] --- we use the \emph{average energy} of each Floquet
eigenstate,
\begin{equation}
\aE_n \;=\; \frac{1}{T}\int_0^T
\langle \phi_n(t)|\,H(t)\,|\phi_n(t)\rangle \, dt ,
\label{eq:ae}
\end{equation}
with $|\phi_n(t)\rangle$ the micromotion of $|u_n\rangle$. For the two-step drive
Eq.~(\ref{eq:UF})  (or any piecewise-constant driving)
the integral is exact and free of any time discretization.
Within the
first half-period $H(t)=\HAF$ and $|\phi_n(t)\rangle=e^{-it\HAF}|u_n\rangle$, so
$\langle\HAF\rangle$ is constant and equal to its $t=0$ value; within the second
half-period $H(t)=\HK$ and $\langle\HK\rangle$ is constant and equal to its value at
$t=T/2$. Hence
\begin{equation}
	\aE_n=\tfrac12\langle u_n|\HAF|u_n\rangle
	+\tfrac12\langle\psi_n|\HK|\psi_n\rangle,
	\qquad |\psi_n\rangle=e^{-i\frac{T}{2}\HAF}|u_n\rangle .
	\label{eq:aeexact}
\end{equation}
The micromotion in fact drops out of Eq.~(\ref{eq:aeexact}). Write
$A=e^{-i\frac{T}{2}\HAF}$ and $B=e^{-i\frac{T}{2}\HK}$, so that $\UF=BA$ and
$|\psi_n\rangle=A|u_n\rangle$. The eigenvalue equation $BA|u_n\rangle
=\lambda_n|u_n\rangle$ gives $A|u_n\rangle=\lambda_n B^\dagger|u_n\rangle$, whence,
using $|\lambda_n|=1$ and $[B,\HK]=0$,
$\langle\psi_n|\HK|\psi_n\rangle=\langle u_n|A^\dagger\HK A|u_n\rangle
=\langle u_n|B\,\HK\,B^\dagger|u_n\rangle=\langle u_n|\HK|u_n\rangle$, so that
\begin{equation}
	\aE_n=\langle u_n|\overline{H}|u_n\rangle
	\label{eq:aeH}
\end{equation}
\emph{exactly}, at every $T$: the average energy is the static Hamiltonian evaluated
in the Floquet eigenstate. The cancellation is specific to the two-step drive: in
the four-step drive of Ref.~\cite{detuning} only the first and last steps collapse,
leaving the rotated operator of its Eq.~(19). As $T\to0$, $|u_n\rangle$ becomes
an eigenstate of $\overline{H}$ and $\aE_n\to\varepsilon_n$. Note that with the
convention (\ref{eq:UF}) the natural static reference for both $\varepsilon_n$ and
$\aE_n$ is $H/2$, so all static gaps are compared after rescaling by $1/2$.
$\aE_n$ is not folded, and therefore \emph{orders} the Floquet spectrum for any $T$.

\begin{figure}[t]
\centering
\includegraphics[width=0.85\textwidth]{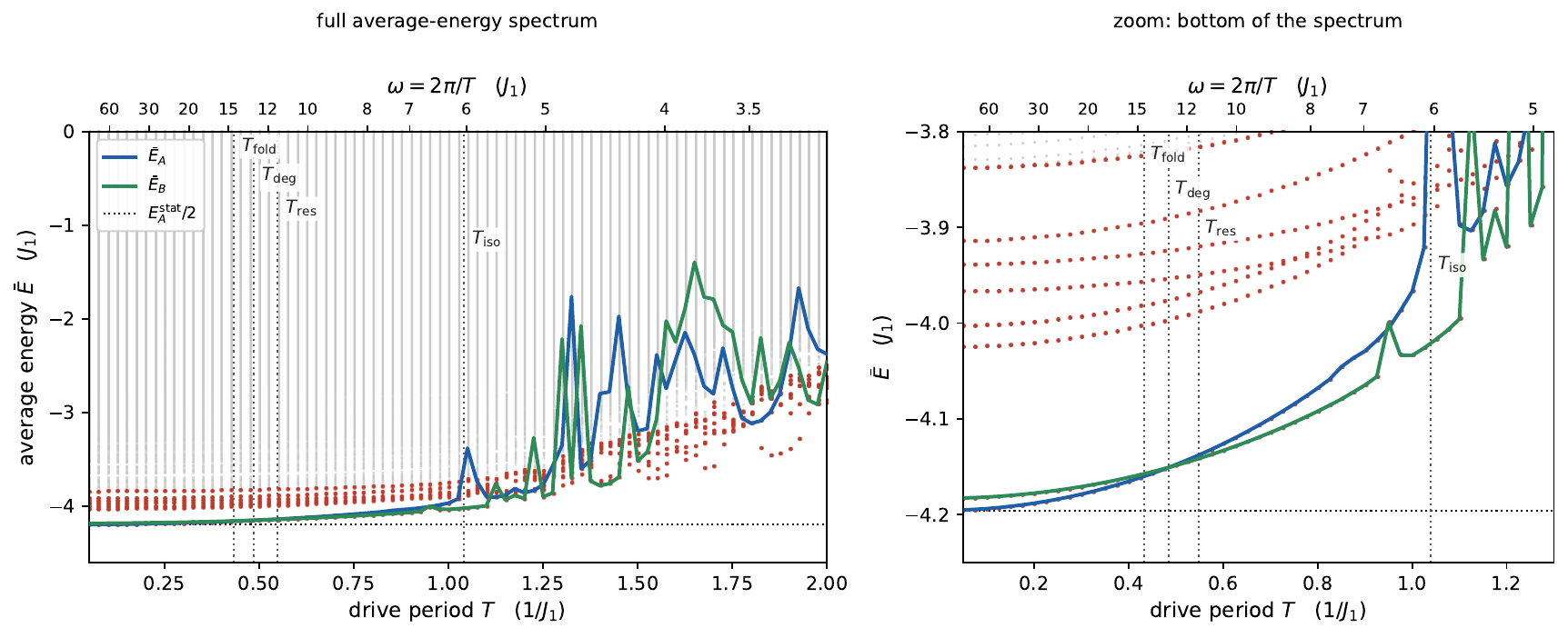}
\caption{Average-energy spectrum versus $T$ (upper axis: $\omega=2\pi/T$).
Grey: all $12870$ levels; red: the lowest eight; blue/green: the topological
doublet followed by overlap. The key periods $T_{\mathrm{fold}}$,
$T_{\mathrm{deg}}$, $T_{\mathrm{res}}$, $T_{\mathrm{iso}}$ are marked. Right
panel: zoom on the bottom of the spectrum, $\bar E\in[-4.25,-3.80]$, where the
doublet is resolved from the red band of low-lying levels. The doublet remains
the isolated bottom of the average-energy spectrum far beyond
$T_{\mathrm{res}}$, up to $T_{\mathrm{iso}}\simeq1.04$.}
\label{fig:aespec}
\end{figure}

\begin{figure}[t]
\centering
\includegraphics[width=0.95\textwidth]{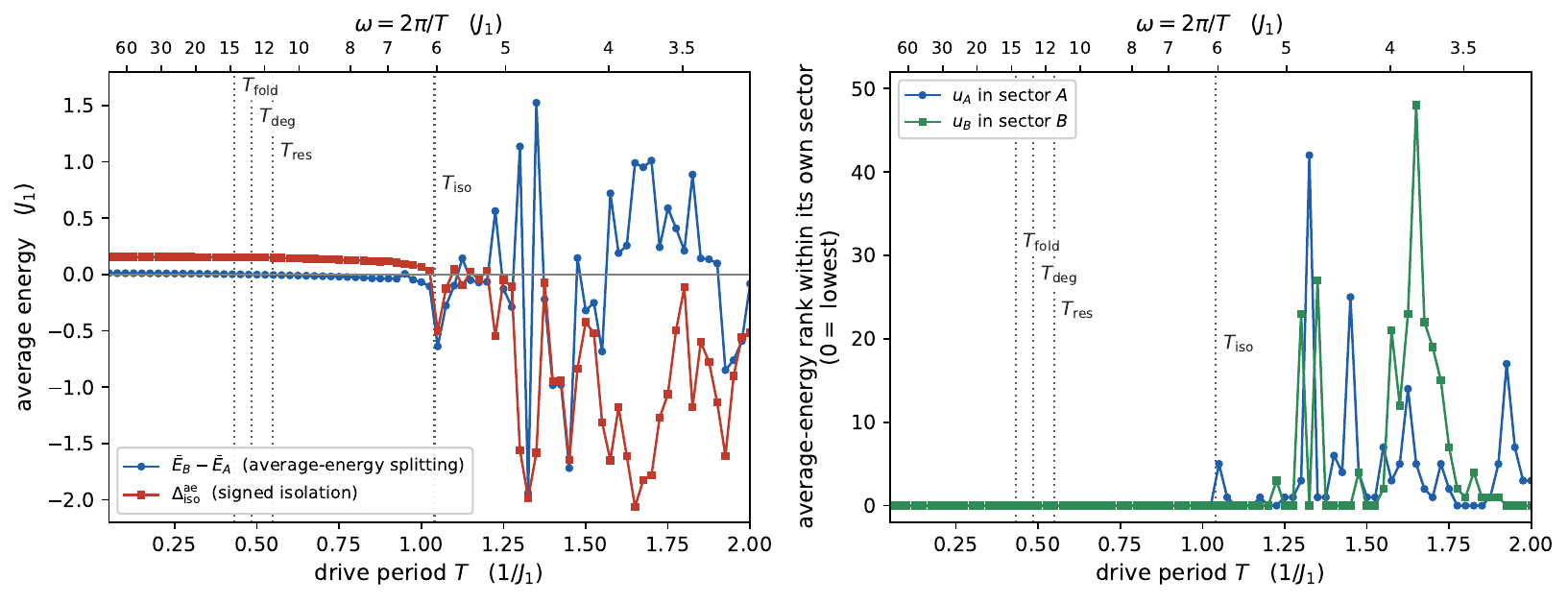}
\caption{Left: doublet splitting and signed isolation in average energy
(negative if an intruder lies below the doublet top), versus $T$ (upper axis:
$\omega=2\pi/T$). Right: rank of each doublet state in the average-energy
ordering of its own symmetry sector ($0=$ lowest). Both members keep rank $0$
through $T_{\mathrm{fold}}$ and $T_{\mathrm{res}}$ and first leave it at
$T_{\mathrm{iso}}$.}
\label{fig:aedoublet}
\end{figure}

The result (Figs.~\ref{fig:aespec},~\ref{fig:aedoublet}) is striking. In average
energy the doublet remains the \emph{absolute bottom} of the whole spectrum ---
rank $0$ in both its sectors, with strictly positive isolation --- up to
\begin{equation}
T_{\mathrm{iso}} \simeq 1.04 ,\qquad
\omega^*\equiv 2\pi/T_{\mathrm{iso}} \simeq 6\,J_1 ,
\label{eq:Tiso}
\end{equation}
i.e.\ almost twice the period $T_{\mathrm{res}}$ at which the quasienergy-circle
diagnostic collapses. Since $T_{\mathrm{iso}}$ is read off a scan grid of step
$\Delta T=0.025$, $\omega^*$ is fixed only to $\simeq0.1\,J_1$ and is to be read
throughout as a local energy scale, not as a sharply quantitative threshold. The comb of quasienergy resonances at
$T=0.55,\,0.60,\,0.65,\dots$ leaves \emph{no visible trace} in $\aE$: the folded
states cross the doublet in quasienergy but essentially do not hybridize with it,
i.e.\ the avoided crossings are parametrically narrow. The first crossing that
actually disturbs the doublet is a strong $A$-sector resonance at $T\simeq1.05$, and
a genuine destruction of the doublet (rank jumps, large negative isolation) only
occurs for $T\gtrsim1.25$.

Two further features of the average-energy analysis are worth noting.
(i) The doublet heats extremely slowly: $\aE_A$ rises from $-4.196$ ($T\to0$) to
$-4.006$ at $T=0.95$ and $-3.967$ at $T=1$, i.e.\ by less than $2\%$ of $W$.
(ii) The average-energy splitting $\aE_B-\aE_A$ changes sign at
\begin{equation}
T_{\mathrm{deg}} = 0.485 \qquad (\omega \simeq 13.0\,J_1),
\end{equation}
an \emph{exact average-energy degeneracy of the topological doublet}, located inside
the folded-but-stable window $[T_{\mathrm{fold}},T_{\mathrm{res}}]$. The drive thus
achieves at $T_{\mathrm{deg}}$ what the static model at this parameter point cannot:
an exactly degenerate, isolated, cold topological doublet --- arguably the optimal
working point of the protocol.

\section{$D=3$ chiral PEPS description of the Floquet doublet}
\label{sec:peps}

A complementary, wavefunction-based diagnostic is provided by the variational
$D=3$ chiral PEPS description of the CSL (Appendix~\ref{app:static}). For the $A$
member of the doublet the ansatz is the uniform PEPS built from the chiral family
of $SU(2)$-symmetric site tensors
${\mathcal A}(a_1,a_2,a_3)=a_1A_1^{(a)}+a_2A_1^{(b)}+ia_3A_2$, first constructed in
Ref.~\cite{poilblanc2015} and classified in Ref.~\cite{mambrini2016}; the $B$ member is reached by threading a $\mathbb{Z}_2$ gauge
string $g=\mathrm{diag}(-1,-1,+1)$ along a non-contractible cycle and
antisymmetrizing the two orientations, $\psi_x-\psi_y$ (an exact $S=0$,
$\mathbf{k}=0$, $C_4=-1$ state). For each period $T$ we optimize the tensor
coefficients against the \emph{Floquet} doublet eigenstates $u_{A,B}(T)$ (taken at
the beginning of the period, i.e.\ at the start of the Heisenberg half-step).
Table~\ref{tab:peps} shows the result at representative periods, including the
average-energy degeneracy point $T_{\mathrm{deg}}=0.485$ of Sec.~\ref{sec:ae}.

\begin{table}[t]
\centering
\begin{tabular}{lcccc}
\toprule
$T$ & $|\langle u_A|\mathrm{PEPS}\rangle|^2$ & $(a_1,a_2,a_3)_A$ &
$|\langle u_B|\psi_x-\psi_y\rangle|^2$ & $(a_1,a_2,a_3)_B$ \\
\midrule
$0$ (static)             & $0.9168$ & $(0.809,0.442,-0.387)$ & $0.7822$ & $(0.791,0.421,-0.443)$ \\
$0.25$                   & $0.9060$ & $(0.810,0.442,-0.386)$ & $0.7779$ & $(0.793,0.417,-0.444)$ \\
$0.485=T_{\mathrm{deg}}$ & $0.8762$ & $(0.810,0.443,-0.383)$ & $0.7662$ & $(0.797,0.408,-0.446)$ \\
$0.75$                   & $0.8200$ & $(0.812,0.446,-0.377)$ & $0.7442$ & $(0.803,0.389,-0.451)$ \\
$1.0$                    & $0.7430$ & $(0.814,0.452,-0.365)$ & $0.7147$ & $(0.812,0.365,-0.456)$ \\
\bottomrule
\end{tabular}
\caption{Optimal squared overlaps of the $D=3$ chiral PEPS ans\"atze with the
Floquet doublet eigenstates versus the drive period, with the corresponding
normalized tensor coefficients. The $T=0$ row reproduces the static benchmarks of
Appendix~\ref{app:static}.}
\label{tab:peps}
\end{table}

Three observations stand out. First, the PEPS description degrades \emph{smoothly}
with $T$ and shows no anomaly whatsoever at the folding scales
$T_{\mathrm{fold}}=0.432$ or $T_{\mathrm{res}}=0.548$: at the working point
$T_{\mathrm{deg}}$, deep in the folded regime, the driven doublet is still described
at the $0.88$ ($A$) and $0.77$ ($B$) level, only a few percent below the static
values. Second, the optimal tensors barely move over the whole range
$T\in[0,1]$ --- the chiral-tensor content of the state is essentially frozen at its
static value. Third, the decrease is quantitatively accounted for by the rotation of
the Floquet state away from the static one: to good accuracy
$|\langle u_A|\mathrm{PEPS}\rangle|^2 \simeq
|\langle\psi_A|\mathrm{PEPS}\rangle|^2\,|\langle\psi_A|u_A\rangle|^2$
(e.g.\ $0.9168\times0.956=0.876$ at $T_{\mathrm{deg}}$, versus $0.8762$ measured),
i.e.\ the loss reflects the micromotion dressing of the Floquet eigenstate by
components orthogonal to the static ground state, \emph{not} a failure of the CSL
tensor manifold.

This last statement can be made sharp, and it also settles how much the reported
overlaps depend on the choice of reference time inside the period. Consider the
time-symmetric version of the protocol,
$U_F^{\mathrm{sym}}(T)=e^{-i\frac{T}{4}\HAF}e^{-i\frac{T}{2}\HK}
e^{-i\frac{T}{4}\HAF}$. It is \emph{unitarily equivalent} to
Eq.~(\ref{eq:UF}), $U_F^{\mathrm{sym}}=X\,\UF X^{-1}$ with
$X=e^{-i\frac{T}{4}\HAF}$, so its quasienergies are identical; and the average
energy (\ref{eq:ae}), being an integral over a full period, is likewise
unchanged --- we have verified both numerically to machine precision. Every
threshold of this paper, $T_{\mathrm{fold}}$, $T_{\mathrm{res}}$,
$T_{\mathrm{deg}}$ and $T_{\mathrm{iso}}$, is therefore \emph{exactly}
independent of that choice. What does change are the overlaps, which are
explicitly reference-time dependent --- and they change a great deal: measured
at the symmetric point, $|\langle\psi_A|u_A\rangle|^2=0.9918$ at $T=1$ and
$0.9701$ at $T=1.25$, against $0.8189$ and $0.7093$ for the convention of
Eq.~(\ref{eq:UF}). Almost the whole overlap deficit of Table~\ref{tab:peps} is
thus removable by a shift of the reference time, which confirms directly that it
is micromotion dressing and not a change of the underlying state. Together with the average-energy analysis, this supports the
picture that the prethermal Floquet Hamiltonian hosts the same CSL phase as the
static model, with an essentially unchanged local tensor description.

\begin{figure}[t]
\centering
\includegraphics[width=0.95\textwidth]{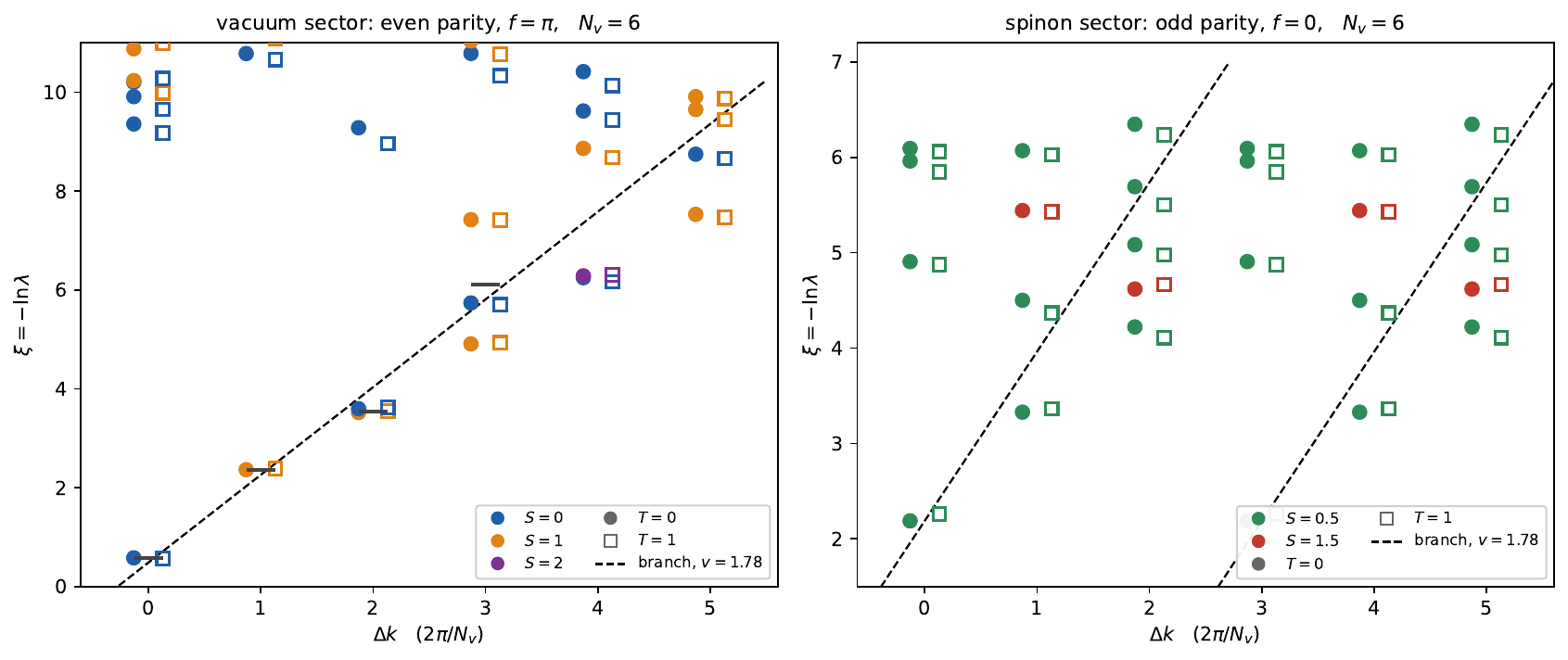}
\caption{Spin-resolved entanglement spectrum of the optimal chiral PEPS on an
infinite cylinder of circumference $N_v=6$, in the physical vacuum sector
[even parity, flux $f{=}\pi$; left] and spinon sector [odd parity, $f{=}0$;
right], for the static ($T{=}0$, filled circles) and Floquet-optimized
($T{=}1$, open squares) tensors, versus $\Delta k$, the momentum measured
from the tower base along the chirality direction. Colors encode the total
spin $S$ of each multiplet. Dashed lines: a \emph{single} straight chiral
branch of slope $v$, obtained by a least-squares fit through the
degeneracy-weighted barycenters of the first four even tower levels, i.e.\ of
the $SU(2)_1$ vacuum content $(0);(1);(1){+}(0);2(1){+}(0)$ at
$\Delta k=0,1,2,3$ (short horizontal dashes in the left panel mark those four
barycenters, at $\xi=0.580,\,2.364,\,3.541,\,6.105$ for $T{=}0$). The fit gives
an edge velocity $v(T{=}0)=1.775\pm0.187$ against $v(T{=}1)=1.781\pm0.182$,
i.e.\ unchanged at the $0.3\%$ level, well inside the fit uncertainty. The
\emph{same} slope --- identical in the two panels, as required for the
$SU(2)_1$ chiral edge --- is then anchored at the two degenerate spinon bases
$\Delta k=0,\pi$ (right). Both panels realize the expected $SU(2)_1$ tower
contents (see text).}
\label{fig:es}
\end{figure}


A sharper fingerprint of the topological order is the entanglement spectrum (ES)
of the half-infinite cylinder. As first discovered by Li and Haldane~\cite{LiHaldane} the ES is in one-to-one correspondance with the physical chiral edge spectrum (labeled by the edge momentum). It can be computed from the PEPS transfer operator following
Ref.~\cite{psa2016} (see Appendix~\ref{app:es} for details).  
Figure~\ref{fig:es} compares the entanglement spectra of the $T{=}0$ and $T{=}1$ optimal
tensors at $N_v=6$ in the physical sectors [$(\mathrm{even},\pi)$ and
$(\mathrm{odd},0)$]. 
The spectra are plotted versus $\Delta k$, the momentum
measured from the tower base along the chirality direction
(the base momenta of all towers are identical).
Both display the hallmarks of an $SU(2)_1$ WZW chiral
edge \cite{psa2016}: (i) all even-sector levels carry \emph{integer} spin and
all odd-sector levels \emph{half-integer} spin, with exact $2S{+}1$
multiplicities; (ii) the even spectrum is a \emph{textbook chiral vacuum tower},
$(0);(1);(1){+}(0);2(1){+}(0);(2){+}\dots$ at successive $\Delta k$, dispersing
in a single direction, with the first $S{=}2$ multiplet appearing exactly at
$\Delta k=4$ as required by the $SU(2)_1$ vacuum character; (iii) the odd
spectrum realizes the $j{=}\tfrac12$ tower
$(\tfrac12);(\tfrac12);(\tfrac12){+}(\tfrac32);2(\tfrac12){+}(\tfrac32);\dots$,
with a \emph{unique} $S{=}\tfrac12$ level at $\Delta k=1$ and the first
$(\tfrac32)$ appearing at $\Delta k=2$, every level doubled at momenta $k$ and
$k+\pi$ (the semionic double degeneracy of the spinon sector).

The salient result is the $T{=}0$ vs $T{=}1$ comparison: the two spectra
coincide multiplet by multiplet --- identical spin content, momenta and
degeneracy patterns, with level shifts of a few percent at most (e.g.\ the
vacuum level moves from $\xi=0.58$ to $0.56$, the first current triplet from
$2.36$ to $2.39$). The chiral branch itself is quantitatively unchanged: fitting
its slope on the vacuum tower (Fig.~\ref{fig:es}) gives an edge velocity
$v=1.775\pm0.187$ at $T{=}0$ against $1.781\pm0.182$ at $T{=}1$, a shift of
$0.3\%$. The $SU(2)_1$ chiral edge of the CSL is thus fully intact in the driven
state.

\section{Stroboscopic dynamics and heating}
\label{sec:strobe}

\begin{figure}[t]
\centering
\includegraphics[width=0.9\textwidth]{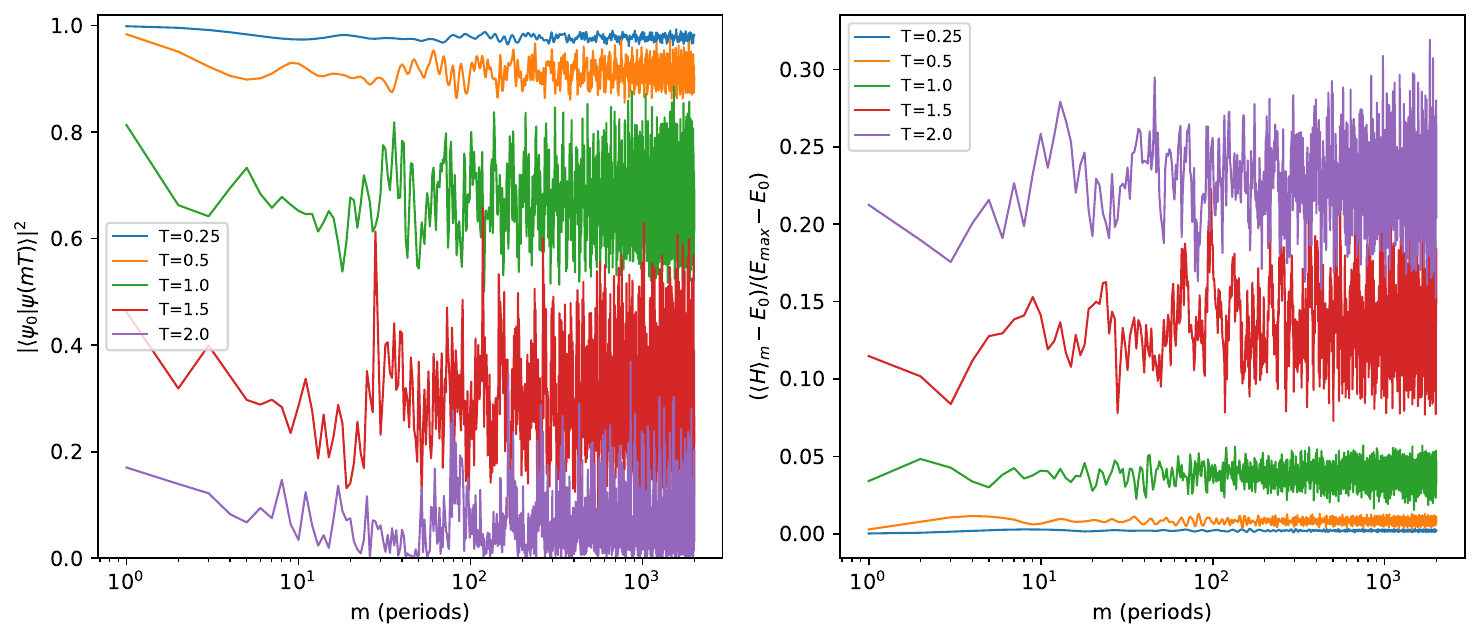}
\caption{Stroboscopic evolution $|\psi(mT)\rangle=\UF^m|\psi_A^{\mathrm{stat}}\rangle$
up to $m=2000$ periods, for several drive periods $T$. Left: fidelity to the initial
state. Right: absorbed energy normalized to the full many-body bandwidth.}
\label{fig:strobe}
\end{figure}

Starting from the static ground state $\psi_0=\psi_A$ we have evolved stroboscopically the state over
$2000$ periods. The fidelity to the initial state and the absorbed energy $(\big<H\big>_m -E_0)/W$
(normalized to the full many-body bandwidth) are shown in Fig.~\ref{fig:strobe}, left and right panels respectively, for various values of $T$. At $T=0.25$ and $T=0.5$ we see fidelity
plateaus at $0.98$ and $0.90$ with less than $1\%$ of the bandwidth absorbed ---
no heating on this timescale, including at $T=0.5$ which lies \emph{beyond the folding
onset}. At $T=1.0$ the plateau is $0.65$--$0.7$ with $\sim4\%$ absorbed; only for
$T\gtrsim1.5$ does the system enter a heating regime (fidelity $0.2$--$0.3$, $13$--$23\%$
absorbed and growing). The heating onset coincides with the destruction of the
doublet in the average-energy ordering, $T\gtrsim1.25$, and \emph{not} with
$T_{\mathrm{res}}$.

Two controls on this comparison are worth recording. First, at fixed $m$ the
elapsed physical time $mT$ varies by a factor of eight across the range of $T$
used, so the runs are not directly comparable as they stand. Repeating the
measurement at a \emph{fixed} physical time $J_1t=500$ changes nothing
qualitative --- the absorbed energy (in units of $W_H$) reads
$0.21\%,\ 0.93\%,\ 3.30\%,\ 14.98\%,\ 18.46\%$ at
$T=0.25,\,0.5,\,1.0,\,1.5,\,2.0$, against
$0.21\%,\ 0.80\%,\ 3.84\%,\ 14.87\%,\ 20.43\%$ at fixed $m=2000$ --- and if
anything sharpens the conclusion, since the low-frequency runs are then the
longest in physical time. Second, nothing depends on which member of the doublet
is prepared: starting from $\psi_B$ instead of $\psi_A$ gives systematically
\emph{better} numbers at every period (at $T=1$, $\max_n|c_n|^2=0.876$ versus
$0.819$, fidelity $0.82$ versus $0.69$, and $\Delta E_{\mathrm{DE}}=2.5\%$
versus $3.8\%$), while a random state in the same sector has
$\max_n|c_n|^2\simeq0.02$ and an inverse participation ratio above $100$.
Since $A$ and $B$ belong to different symmetry sectors, $\UF$ cannot mix them,
so an arbitrary superposition of the two --- the state of the topological qubit
--- inherits the same stability, its relative phase winding at the rate
$\Delta^F_{AB}$, which the drive \emph{reduces} (Sec.~\ref{sec:quasi}).

\begin{figure}[t]
\centering
\includegraphics[width=0.62\textwidth]{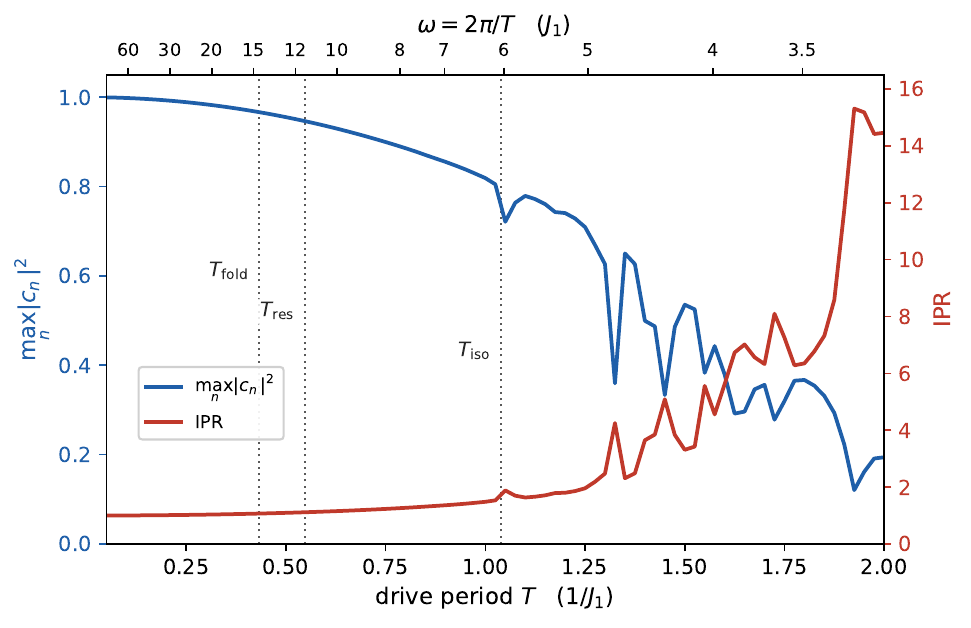}
\caption{Participation of the initial state $\psi_0=\psi_A^{\mathrm{stat}}$ over
the Floquet eigenbasis of sector $A$ versus the drive period $T$ (upper axis:
$\omega=2\pi/T$): the largest weight $\max_n|c_n|^2$ on a single Floquet
eigenstate (left scale, blue) and the inverse participation ratio (right scale,
red). The initial state remains essentially a single Floquet eigenstate through
both folding scales, and delocalizes only beyond $T_{\mathrm{iso}}$.}
\label{fig:part}
\end{figure}

\begin{figure}[t]
\centering
\includegraphics[width=0.95\textwidth]{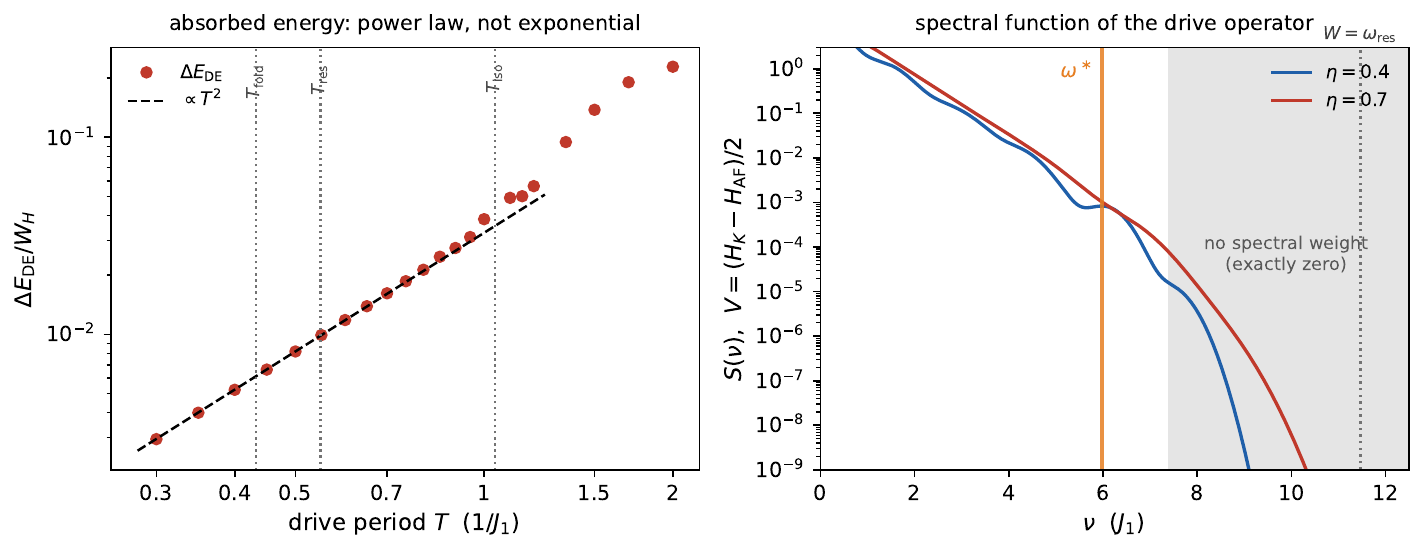}
\caption{Left: diagonal-ensemble energy absorption $\Delta E_{\mathrm{DE}}$,
Eq.~(\ref{eq:dEDE}), normalized to $W_H=2W$, versus the drive period on a
double-logarithmic scale. The dashed line is the $T^2$ law of
Eq.~(\ref{eq:T2}); the data follow it through both folding scales
$T_{\mathrm{fold}}$ and $T_{\mathrm{res}}$ without any feature, and depart from
it only near $T_{\mathrm{iso}}$. Right: spectral function $S(\nu)$,
Eq.~(\ref{eq:Snu}), of the drive operator $V=(\HK-\HAF)/2$ in the static eigenbasis, Gaussian-broadened with two
widths $\eta$ to show that the shape is not an artefact of the broadening. The
shaded region carries no spectral weight at all (below $10^{-30}$); the
threshold $\omega^*=2\pi/T_{\mathrm{iso}}$ measured by the other diagnostics is
marked, and coincides with the edge of the support. The extensive scale $W=\omega_{\mathrm{res}}$ is
shown for comparison and plays no role.}
\label{fig:heat}
\end{figure}

A comment is in order on what ``heating'' means on a finite cluster, and how it
is quantified. Since the Floquet spectrum of a finite system is discrete, the
stroboscopic dynamics is quasi-periodic: after a transient, all observables
oscillate around the predictions of the \emph{diagonal ensemble} in the
Floquet eigenbasis, $\rho_{\mathrm{DE}}=\sum_n |c_n|^2 |u_n\rangle\langle
u_n|$ with $c_n=\langle u_n|\psi_0\rangle$; a finite cluster never flows all
the way to infinite temperature unless the drive hybridizes the initial state
with the bulk of the spectrum. How much it does so is shown in
Fig.~\ref{fig:part}: at high frequency the initial state coincides almost entirely
with a single Floquet eigenstate --- the doublet-$A$ member --- so its largest weight
$\max_n|c_n|^2$ stays close to unity and the inverse participation ratio remains
minimal; only for $T\gtrsim1.25$, coincident with the loss of the doublet, does
$\psi_0$ delocalize over many Floquet eigenstates ($\max_n|c_n|^2$ collapsing and the
IPR growing). The sharp, time-independent diagnostic of the resulting energy
offset is therefore the diagonal-ensemble absorption
\begin{equation}
	\Delta E_{\mathrm{DE}}={\rm Tr}\{\rho_{\mathrm{DE}}(H-E_0)\}=\sum_n |c_n|^2\langle u_n|H|u_n\rangle-E_0,
	\label{eq:dEDE}
\end{equation}
where $H=\HAF+\HK$ is the \emph{static} Hamiltonian and $E_0=-8.3913$ its
ground-state energy in sector $A$; we normalize it by the spread $W_H=2W=22.94$
of that same $H$, so that the percentages quoted below are fractions of $W_H$. Since
$\langle u_n|H|u_n\rangle=2\aE_n$ by Eq.~(\ref{eq:aeH}), $\Delta E_{\mathrm{DE}}$ and
the average-energy ordering are built from the same per-eigenstate quantity ---
weighted by $|c_n|^2$ here, ranked there --- so the coincidence of the heating onset
with $T_{\mathrm{iso}}$ is not accidental.

The result is shown in the left panel of Fig.~\ref{fig:heat}. Over the whole
range explored, $\Delta E_{\mathrm{DE}}$ is accurately a \emph{power law} in the
period,
\begin{equation}
\Delta E_{\mathrm{DE}}/W_H \;\simeq\; 0.0328\;T^2
\qquad (T\lesssim1),
\label{eq:T2}
\end{equation}
the ratio $\Delta E_{\mathrm{DE}}/T^2$ being constant to four significant figures
between $T=0.30$ and $T=0.70$. (A single exponential $e^{-c\omega/J_1}$ fitted
to the same data describes it poorly: the decay constant drifts from $0.52$ to
$0.18$ across sub-windows, with residuals three to ten times larger.)

The $T^2$ law is exactly what should be expected, and it identifies
$\Delta E_{\mathrm{DE}}$ as \emph{dressing rather than heating}. Being a
diagonal-ensemble quantity, $\Delta E_{\mathrm{DE}}$ is time independent: it
measures how far $\psi_0$ is from being a single Floquet eigenstate, not a rate.
Since $H_F=\overline{H}+O(T)$ by the Magnus expansion, $|u_A\rangle=|\psi_A\rangle
+O(T)$; and because $\psi_A$ is the ground state of $H$ the term linear in $T$ in
$\langle u_A|H|u_A\rangle-E_0$ vanishes, leaving $O(T^2)$. Decomposing
Eq.~(\ref{eq:dEDE}) into the single-eigenstate part $\langle u_A|H|u_A\rangle-E_0$
and the delocalization part $\sum_{n\neq A}|c_n|^2(\langle u_n|H|u_n\rangle-
\langle u_A|H|u_A\rangle)$ confirms this: both scale as $T^{2.0}$.
What survives, and is the point of the figure, is that $\Delta E_{\mathrm{DE}}$
is small and perfectly smooth --- with \emph{no feature whatsoever} at
$\omega_{\mathrm{fold}}$ or $\omega_{\mathrm{res}}$ --- rising above the $T^2$ law
only near $T_{\mathrm{iso}}$.

The genuinely exponential object of the prethermalization bounds
\cite{mori2016,kuwahara2016,abanin2017} is the heating \emph{rate}, set in linear
response by the spectral function of the drive operator $V=(\HK-\HAF)/2$
\cite{mallayya2019} --- writing the two-step drive as
$H(t)=\overline{H}+V\!f(t)$ with $f(t)=\pm1$ the square wave that is $+1$ on the
chiral half-step, $V$ is simply its modulation amplitude, and its overall sign
is immaterial since only $|\langle\alpha|V|\psi_0\rangle|^2$ enters. The object
that controls the rate, evaluated at the drive harmonics, is
\begin{equation}
S(\nu)=\sum_\alpha\big|\langle\alpha|V|\psi_0\rangle\big|^2\,
\delta\big(\tfrac12(E_\alpha-E_0)-\nu\big),
\label{eq:Snu}
\end{equation}
where $\{|\alpha\rangle,E_\alpha\}$ are the eigenstates and eigenvalues of the
\emph{static} Hamiltonian $H$, and the factor $\tfrac12$ places $\nu$ on the same
scale as the quasienergies and as $\omega$ (convention $\overline{H}=H/2$). On the $4\times4$ torus this quantity cannot be used to extract
an exponential --- but it delivers something better. As the right panel of
Fig.~\ref{fig:heat} shows, $S(\nu)$ has \emph{compact support}: $99.9\%$ of its
weight lies below $\nu=4.5\,J_1$ and the last transition with non-negligible
weight sits at $\nu=7.4\,J_1$, above which the weight is zero to machine
precision. Absorption at $\omega$ beyond that window is therefore not
exponentially suppressed on this cluster, it is kinematically forbidden. The
essential point is that this window is set by \emph{locality}: $V$ is a sum of
local terms, so $V|\psi_0\rangle$ can only reach states within an intensive
energy of the ground state, however large the system. And the edge of the window
coincides with the threshold measured by every other diagnostic:
$\omega^*\simeq5$--$6\,J_1$ is precisely where the spectral weight of the drive
operator dies. Whether the residual tail \emph{beyond} the support is
exponentially small in $\omega$, as the rigorous bounds require, is a question
that needs the larger clusters of prediction~(i) below.

In the thermodynamic limit the same physics appears as a two-step relaxation:
fast approach to a prethermal state governed by $H_F^{\mathrm{eff}}$, followed
by exponentially slow heating, as established numerically for generic driven
lattice models \cite{dalessio2014,lazarides2014,ponte2015,bukov2015,machado2020}
and observed experimentally in driven Bose--Hubbard systems \cite{rubio2020}
and dipolar spin ensembles \cite{beatrez2021}. Standard quantitative tests,
all applicable to larger clusters or future experiments on this model,
include: (i) the absorbed energy versus $\omega$ at fixed evolution time (as
in Fig.~\ref{fig:heat}) and its finite-size scaling; (ii) direct evaluation
of the Fermi-golden-rule heating rate from the spectral function of
$\HK-\HAF$ \cite{mallayya2019}; (iii) the crossover of the quasienergy
level statistics of $\UF$ towards the circular-ensemble result when heating
sets in \cite{dalessio2014}; and (iv) the growth rate of the entanglement
entropy under $\UF^m$. Diagnostic (iii) deserves a comment, since the full set of
quasienergies is in hand at every $T$. It requires resolving \emph{every}
symmetry, and $[\UF,\mathbf{S}^2]=0$ here, so the momentum/$C_4$ sectors must be
decomposed further into total-spin blocks; unresolved, they return the Poisson
value $\langle r\rangle\simeq0.39$ at all $T$ irrespective of ergodicity. Once
resolved, the largest available block on the $4\times4$ torus holds $101$ levels
($\mathbf{k}=(\pi,\pi)$, $S=0$; the $\mathbf{k}=0$ blocks hold $42$ and $39$),
giving a statistical uncertainty on $\langle r\rangle$ of about $0.10$ --- half
of the entire Poisson-to-circular-ensemble range $0.386\to0.600$. The test is
therefore not conclusive at this size, and we defer it to the larger clusters of
prediction~(i). On our cluster, diagnostic (i) and the average-energy
ordering of Sec.~\ref{sec:ae} give a consistent heating threshold
$\omega^*\simeq5$--$6\,J_1$, far below $\omega_{\mathrm{res}}$.

\section{A local order parameter: the scalar chirality}
\label{sec:chir}

\begin{figure}[t]
\centering
\includegraphics[width=0.95\textwidth]{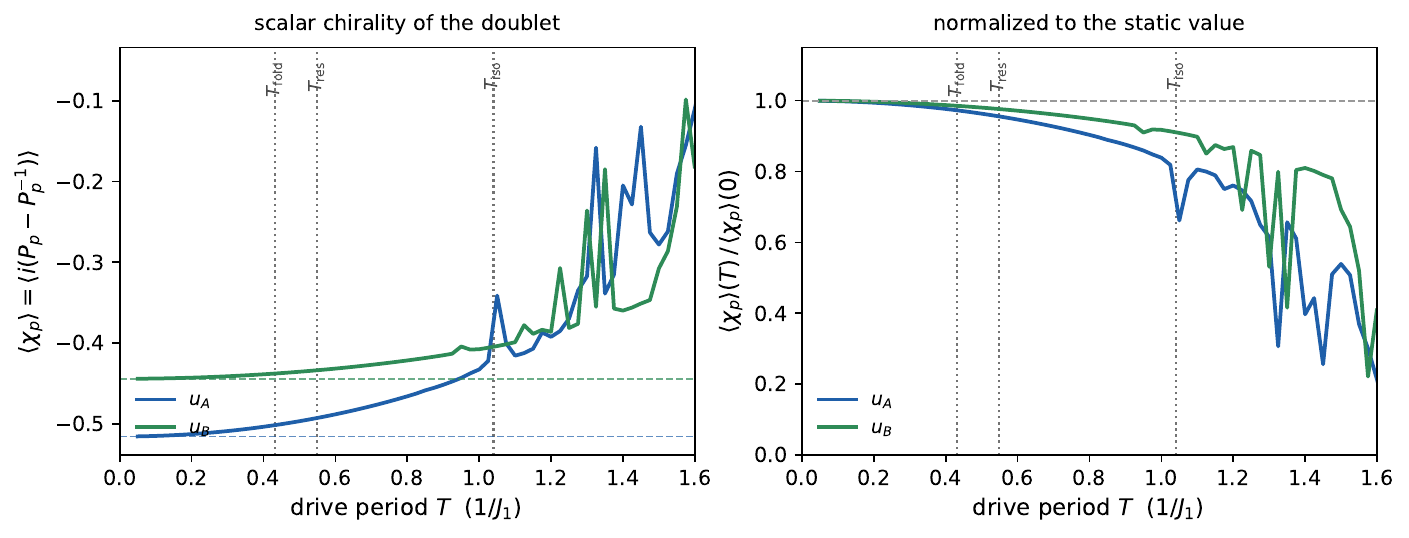}
\caption{Scalar chirality per plaquette
$\langle\chi_p\rangle=\langle i(P_p-P_p^{-1})\rangle$ of the two Floquet
doublet eigenstates versus the drive period. Left: absolute values, with the
static values as dashed lines. Right: normalized to the static value. The
folding scales $T_{\mathrm{fold}}$ and $T_{\mathrm{res}}$, and the
average-energy threshold $T_{\mathrm{iso}}$, are marked. The order parameter is
flat across both folding scales and collapses only beyond $T_{\mathrm{iso}}$.}
\label{fig:chir}
\end{figure}

All the diagnostics used so far are spectral, variational or entanglement based.
It is worth closing with a \emph{local order parameter}, since the claim of this
work is precisely that local physics, not the global bandwidth, controls the
stability. The natural choice is the scalar chirality per plaquette,
$\langle\chi_p\rangle=\langle i(P_p-P_p^{-1})\rangle$, which is the order
parameter conjugate to the term that breaks time reversal, and which is
measurable in principle on any of the platforms of Appendix~\ref{app:pulses}.

Figure~\ref{fig:chir} shows it for both members of the driven doublet. In the
static limit $\langle\chi_p\rangle=-0.5159$ ($A$) and $-0.4444$ ($B$).
Under the drive it is essentially untouched through the folding region: at
$T_{\mathrm{fold}}$ it retains $97.4\%$ ($A$) and $98.6\%$ ($B$) of its static
value, and at $T_{\mathrm{res}}$ still $95.5\%$ and $97.6\%$. It then decreases
smoothly --- $84\%$ and $92\%$ at $T=1$, $72\%$ and $86\%$ at $T=1.25$ ---
and collapses only beyond $T_{\mathrm{iso}}$, reaching $40\%$ ($A$) at
$T=1.4$ with strong period-to-period fluctuations that signal the loss of a
well-defined doublet. A local observable that is blind to $\omega_{\mathrm{fold}}$
and $\omega_{\mathrm{res}}$ but tracks $\omega^*$ is, we believe, the most direct
statement of the thesis of this paper.

\section{Discussion: a stable region beyond $\omega_{\mathrm{res}}$ in the
thermodynamic limit?}
\label{sec:discussion}

The four independent diagnostics --- quasienergy circle, average energy, PEPS
description, and stroboscopic dynamics --- consistently locate the destruction of the Floquet CSL at
$\omega\simeq5$--$6\,J_1$, far below $\omega_{\mathrm{res}}=W\simeq11.5\,J_1$. This
hierarchy is the central observation of this work, because the two scales behave
very differently with system size.

The scale $\omega_{\mathrm{res}}=W$ is \emph{extensive}: on larger clusters the
many-body spread grows like the number of sites ($N=16$ here) so the criterion ``$\omega$ larger than the unfolded
bandwidth'' is unattainable in the thermodynamic limit --- folding, and the resonances
that come with it, are unavoidable at any fixed frequency. If
$\omega_{\mathrm{res}}$ controlled the stability of the driven CSL, no Floquet CSL
would exist in the thermodynamic limit.

Our data show, however, that crossing $\omega_{\mathrm{res}}$ is harmless. The folded
states that cross the doublet between $T_{\mathrm{res}}$ and $T_{\mathrm{iso}}$ are
many-body states at energy $\sim W$ above it: hybridizing with them requires the
drive to deposit an extensive amount of energy in a single Floquet cycle, through a
\emph{local} operator ($H$ itself). The corresponding matrix elements are
exponentially small --- this is the physics underlying rigorous prethermalization
bounds, which guarantee heating rates $\Gamma\sim e^{-c\,\omega/J_{\mathrm{loc}}}$
with $J_{\mathrm{loc}}$ a local energy scale
\cite{mori2016,kuwahara2016,abanin2017}. In our finite cluster this exponential
smallness is directly visible: the resonances are sharp in quasienergy but leave no
trace in average energy (Sec.~\ref{sec:ae}) and no trace in the stroboscopic
dynamics over $2000$ periods (Sec.~\ref{sec:strobe}).

What finally kills the doublet, at $\omega\simeq5$--$6\,J_1$, is not the global
spread $W$ but a \emph{local} scale: $\omega$ becomes comparable to a few times the
natural local energies of the problem --- the doublet sits $\sim4J_1$ below the
bottom edge of the excited manifold of $\overline{H}$ (i.e.\ $\sim8J_1$ in the
static $H$), and the relevant plaquette and bond scales of $\overline{H}$ are of
order $J_1$--$4J_1$. So low-order photon processes can
connect it to nearby excited states with $O(1)$ matrix elements. Local scales are
size-independent. On the strength of this argument --- which, we emphasize, is a
physical argument supported by data at a single system size, not a demonstrated
finite-size trend --- we expect the phenomenology observed here to survive in
the thermodynamic limit in the following prethermal form: for
$\omega>\omega^*\!\sim\!\mathrm{few}\,J_1$ (a threshold set by local physics, on the
$4\times4$ torus $\omega^*\simeq6J_1$), the stroboscopic dynamics is governed for
exponentially long times $t_*\sim e^{c\,\omega/J_1}$ by a prethermal Floquet
Hamiltonian $H_F^{\mathrm{eff}}$ which is a local, quasi-static deformation of
$\overline{H}=H/2$ --- and which therefore hosts the same gapped CSL phase, with its
topological doublet, chiral edge physics and quantized response. In this window the
finite-size resonance comb becomes dense but exponentially weak, and coarse-grained
observables are insensitive to it; heating is a parametrically slow leak rather than
a resonant instability. The average energy of Eq.~(\ref{eq:ae}) is the natural
order parameter for this scenario on finite clusters, since it filters out the
measure-zero resonances and exposes the underlying prethermal ordering.

A natural objection is that the static point (\ref{eq:sweetspot}) is fine-tuned
(Appendix~\ref{app:static} shows how delicately), so that $\omega^*$ might be a
property of that point rather than a local scale. It is not. Repeating the
average-energy analysis elsewhere in the CSL pocket gives

\begin{center}
\begin{tabular}{lccccc}
\toprule
$(\theta/\pi,\,K,\,J_2)$ & $\Delta_{AB}$ & $\Delta_{\mathrm{iso}}$ &
$T_{\mathrm{iso}}$ & $\omega^*=2\pi/T_{\mathrm{iso}}$ \\
\midrule
$(0.325,\,0.20,\,0.40)$ \ (sweet spot) & $+0.0252$ & $0.3149$ & $1.050$ & $6.0$\\
$(0.325,\,0.20,\,0.42)$ & $+0.0268$ & $0.3130$ & $1.050$ & $6.0$\\
$(0.325,\,0.22,\,0.40)$ & $-0.0097$ & $0.1930$ & $0.925$ & $6.8$\\
$(0.340,\,0.20,\,0.40)$ & $+0.0818$ & $0.3421$ & $1.050$ & $6.0$\\
\bottomrule
\end{tabular}
\end{center}

\noindent
Across these points the static diagnostics vary enormously --- $\Delta_{AB}$ by a
factor of eight \emph{and} in sign, $\Delta_{\mathrm{iso}}$ by nearly a factor of
two --- while $\omega^*$ is unchanged, within the grid resolution quoted above,
for three of the four points and moves by $13\%$ for the fourth. This is what one expects of a threshold fixed by local energy scales
($J_1$, $K$, $J_2$ and the plaquette scales built from them) rather than by the
fine-tuned near-degeneracy of the doublet, and it is a further indication that
$\omega^*$ will survive the thermodynamic limit in the form conjectured here.

Beyond the general prethermal argument, two observations specific to this model
support the conjecture. First, the drive \emph{improves} the topological degeneracy
before any resonance is reached (Sec.~\ref{sec:quasi}), and provides an exactly
degenerate point $T_{\mathrm{deg}}$ inside the already-folded window --- the working
point is not fine-tuned against folding. Second, the destruction observed at
$\omega\simeq6J_1$ proceeds through a small number of identifiable low-order
resonances rather than through a broad continuum, as expected when the density of
\emph{effectively coupled} states, not of all states, is what matters.

Verifiable predictions follow. (i) On larger clusters ($4\times6$, tilted 20- and
26-site tori) $\omega_{\mathrm{res}}=W$ grows extensively while $T_{\mathrm{iso}}$
--- extracted from the average-energy ordering --- should stay at
$\omega^*\simeq\mathrm{few}\,J_1$, up to weak finite-size drift.
(ii) The absorbed energy per period in the window
$\omega^*<\omega<W$ should scale exponentially in $\omega$, measurable from
stroboscopic runs at fixed $m$. (iii) The avoided-crossing widths of the resonance
comb should shrink exponentially with the number of local rearrangements needed to
match the folded state, accessible from fine $T$-scans around each crossing.
Prediction (i) is the decisive one, and we regard it as the natural next step
rather than as an established result of the present work.

Finally, the scenario is testable not only numerically but experimentally, and more
easily than the fine-tuned Hamiltonian of Eq.~(\ref{eq:sweetspot}) might suggest.
Appendix~\ref{app:larger} shows that on the $26$-site torus the purely chiral point
$\theta=\pi/2$ --- with or without $J_2$ --- hosts a topological doublet better
isolated than the sweet spot itself, so neither the real part of the cyclic term nor
the second-neighbour coupling is essential; and Appendix~\ref{app:pulses} then
assembles $\UF(T)$ from XY-type two-spin gates alone, reproducing the exact
propagator to $1.5\times10^{-5}$ per period. Since the digital circuit carries its
own frequency scale set by the pulse structure, testing stroboscopic stability
beyond the folding frequency \emph{within} such a circuit --- on a Rydberg-atom or
polar-molecule array, or in numerics --- is a concrete route to the physics
discussed here.

\section*{Acknowledgments}

The author designed the scientific framework, formulated the core physical
models, and supervised the entire research process. The physical calculations,
data visualization (figures), and the initial draft of the manuscript were
generated using the AI assistant Claude (model identifier
\texttt{claude-fable-5}, developed by Anthropic), operating
under the explicit, real-time guidance and specific iterative prompts of the
author. To ensure the reliability and reproducibility of the newly generated
codes, they were rigorously benchmarked and validated against several
previously published results by the author. All final outputs, including
mathematical derivations, numerical results, and textual content, were
meticulously verified, corrected, and approved by the author, who assumes
full scientific and ethical responsibility for the integrity of this work.
The complete simulation codes generated and used in this work are publicly
available in the GitHub repository
\url{https://github.com/didierpoilblanc/Floquet-CSL-simulations}.
Also, the author thanks Marin Bukov, Sylvain Capponi and Matthieu  Mambrini for  stimulating discussions.
Appendix~\ref{app:pulses} benefited from valuable discussions with Antoine Browaeys. 

\appendix

\section{Static model: optimization of the $J_1$--$J_2$--$K$ CSL point}
\label{app:static}

This appendix summarizes the earlier static study which fixed the parameter point
(\ref{eq:sweetspot}) used in the main text. We have performed exact diagonalization on the $4\times4$
torus (16 sites, PBC), fully resolved in momentum, $C_4$ and total spin, of
$H=\HAF+\HK$ [Eqs.~(\ref{eq:HAF})--(\ref{eq:HK})]. The CSL diagnostics are the
splitting $\Delta_{AB}=E_B-E_A$ of the topological doublet ($A$, $B$: lowest singlets
at $\mathbf{k}=0$ with $C_4=\pm1$) and its isolation $\Delta_{\mathrm{iso}}$, the gap
between the doublet top and the nearest state in \emph{any} symmetry sector
(negative if an intruder lies below).

\begin{figure}[t]
\centering
\includegraphics[width=\textwidth]{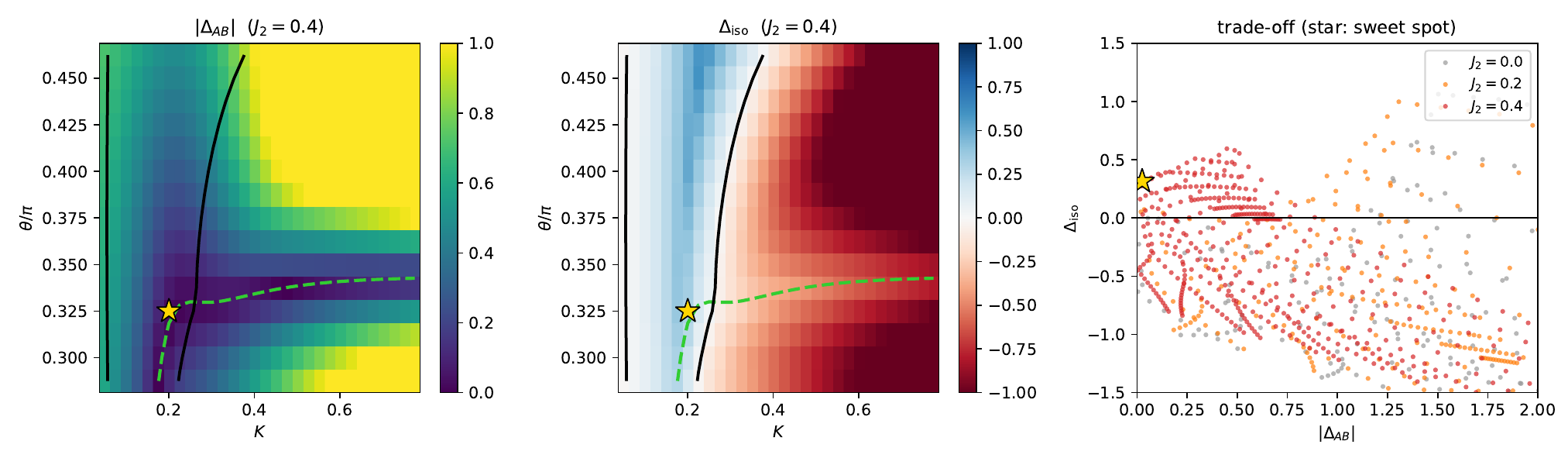}
\caption{Static optimization at $J_2=0.4$ in the $(K,\theta/\pi)$ plane.
Left: doublet splitting $|\Delta_{AB}|$; middle: doublet isolation
$\Delta_{\mathrm{iso}}$ (negative where an intruder lies below the doublet top);
right: trade-off between the two diagnostics for all grid points at
$J_2=0,\,0.2,\,0.4$. The green dashed line is the exact $\Delta_{AB}=0$ crossing
line, the black line the $\Delta_{\mathrm{iso}}=0$ boundary, and the star marks the
selected sweet spot (\ref{eq:sweetspot}), where the crossing line lies inside the
isolated region.}
\label{fig:map}
\end{figure}

\paragraph{Search.} At $J_2=0$ no point is simultaneously quasi-degenerate and
isolated: lowering $\theta$ closes $\Delta_{AB}$, but a $\mathbf{k}=(\pi,\pi)$
singlet collapses faster and becomes the ground state for $\theta/\pi\lesssim0.425$.
The intruder being N\'eel-tower-like, a frustrating $J_2$ pushes it up while barely
affecting the doublet: a modest pocket opens at $J_2=0.2$, and a far better one at
$J_2=0.4$ (Fig.~\ref{fig:map}), flat-optimal in the range $J_2=0.40$--$0.42$, where
the $\Delta_{AB}=0$ crossing line runs through the isolated region. The selected
"sweet spot",
\begin{equation}
(\theta/\pi,\,K,\,J_2)=(0.325,\,0.20,\,0.40):\qquad
\Delta_{AB}=+0.025,\quad \Delta_{\mathrm{iso}}=+0.315,
\end{equation}
has $E_A=-8.391343$, $E_B=-8.366186$, nearest competitor a $\mathbf{k}=(\pi,\pi)$
singlet (all nine sectors checked). The ground state is strongly complex,
$|\sum_i\psi_i^2|/\lVert\psi\rVert^2=0.512$.

\paragraph{$D=3$ chiral PEPS.} Following the construction of Ref.~\cite{poilblanc2015} and the tensor classification of
Ref.~\cite{mambrini2016} a $SU(2)$-symmetric chiral PEPS family
${\mathcal A}=a_1A_1^{(a)}+a_2A_1^{(b)}+ia_3A_2$, $a_i\in \mathbb R$, with virtual space $\tfrac12\oplus0$ (bond dimension $D=3$)
 is constructed  using two (one) site tensors belonging to the $A_1$ ($A_2$) irreducible representation of the $C_{4v}$ point group.
The optimized PEPS reaches its best squared overlap with
$\psi_A$ precisely at the above sweet spot: $0.9168$ (against $0.8974$ at the previously used
$J_2=0$ point $\theta/\pi=0.5$, $K=0.433$), with variational energy $-8.118$
(exact: $-8.391$, i.e.\ a relative variational error of $3.3\%$ --- the scale
against which the overlap degradation of Table~\ref{tab:peps} should be judged). The purely real family ($a_3=0$) degrades there ($0.649$), consistent with
the strongly complex exact state. The $B$ member of the doublet is reached by
threading a $\mathbb{Z}_2$ gauge string along a non-contractible cycle in the two $x$ or $y$ directions:
$\max|\langle\psi_B|\psi_x-\psi_y\rangle|^2=0.782$, while the uniform $B_1+iB_2$
family is gauge-equivalent to $A_1+iA_2$ and has exactly zero overlap with $\psi_B$
on the even torus.

\subsection{Larger clusters: which static point should one target?}
\label{app:larger}

The sweet spot (\ref{eq:sweetspot}) was optimized on the $4\times4$ torus, and two
of its ingredients are awkward from the point of view of a quantum simulator: the
real part $\cos\theta\,(P_p+P_p^{-1})$ of the cyclic term, which a two-body gate set
does not produce naturally (Appendix~\ref{app:pulses}), and the frustrating $J_2$,
which costs extra addressing layers. It is therefore worth asking whether either is
actually necessary --- a question the $4\times4$ cluster alone cannot answer. We
have performed symmetry-resolved Lanczos ED in the $S^z=0$ sector on the
$C_4$-symmetric tori $T_1=(a,b)$, $T_2=(-b,a)$ with $N=a^2+b^2$: $N=16\,(4,0)$,
$20\,(4,2)$ and $26\,(5,1)$. The space group used is
$C_4\ltimes\text{translations}$, $|G|=4N$ --- \emph{without} reflections, which
reverse the plaquette orientation and are not symmetries of $\HK$. Total spin is
obtained from $\langle\mathbf{S}^2_{\mathrm{tot}}\rangle$. The implementation
reproduces the $4\times4$ numbers quoted above exactly ($\Delta_{AB}=0.0252$ versus
$0.025$, $\Delta_{\mathrm{iso}}=0.3149$ versus $0.315$) and was checked against
brute-force full-Hilbert-space diagonalization on $N=8,10$ (agreement to
$2\times10^{-14}$). Besides the sweet spot we consider two purely chiral points at
$\theta=\pi/2$: the ``simple'' model $(J_2,K)=(0,\,0.5)$, and the
$J_1$--$J_2$--$K$ point $(0.40,\,0.25)$.

These two are not arbitrary. Up to an overall normalization they are the standard
reference points of the chiral antiferromagnetic Heisenberg model: the first
coincides with the $J_1$--$\lambda_c$ model $(J_1,J_2,\lambda_c)=(2,0,1)$, and the
second lies close to the $J_1$--$J_2$--$\lambda_c$ model
$(1.78,\,0.84,\,0.375)$, i.e.\ $(J_2,K)/J_1=(0.47,\,0.21)$, both studied with PEPS
in Ref.~\cite{poilblanc2017} and the latter also in Ref.~\cite{hasik2022}. That
second point descends from the local truncation of the Kalmeyer--Laughlin parent
Hamiltonian constructed by Nielsen, Sierra and Cirac \cite{nielsen2013}, in which
the nonlocal parent Hamiltonian is reduced to nearest- and next-nearest-neighbour
Heisenberg couplings plus a scalar-chirality term with position-independent
strengths. Comparing the two exhibits precisely the trend we exploit below:
switching on a frustrating $J_2\simeq0.4$--$0.5\,J_1$ allows the chiral coupling to
be reduced by more than a factor of two, from $K/J_1=0.5$ to $\simeq0.2$, without
losing the CSL. Both points lie inside the CSL region of the $(J_2,K)$ phase
diagram of Ref.~\cite{zhang2024}.

\begin{table}[t]
\centering
\begin{tabular}{clllccc}
\toprule
$N$ & parameters & ground state & partner & $\Delta_{AB}$ & $\Delta_{\mathrm{iso}}$ & $\Delta_{\mathrm{iso}}/\Delta_{AB}$\\
\midrule
16 & sweet spot & $(0,0)\,A$ & $(0,0)\,B$ & 0.0252 & 0.3149 & 12.5\\
20 & sweet spot & $(0,0)\,B$ & $(0,0)\,A$ & 0.0527 & 0.0724 & 1.4\\
26 & sweet spot & $(\pi,\pi)\,A$ & $(0,0)\,E_1$ & 0.0223 & 0.1309 & 5.9\\
\midrule
16 & simple & $(0,0)\,A$ & --- & 1.2133 & 0.0260 & ---\\
20 & simple & $(0,0)\,B$ & $(0,0)\,A$ & 0.4374 & 0.3506 & 0.8\\
26 & simple & $(\pi,\pi)\,A$ & $(0,0)\,E_1$ & 0.0737 & 0.8634 & \textbf{11.7}\\
\midrule
16 & $J_1$--$J_2$--$K$ & $(0,0)\,A$ & $(0,0)\,B$ & 0.6212 & 0.6491 & 1.0\\
20 & $J_1$--$J_2$--$K$ & $(0,0)\,B$ & $(0,0)\,A$ & 0.3910 & 0.5663 & 1.4\\
26 & $J_1$--$J_2$--$K$ & $(\pi,\pi)\,A$ & $(0,0)\,E_1$ & 0.0537 & 0.9198 & \textbf{17.1}\\
\bottomrule
\end{tabular}
\caption{Topological doublet on $C_4$-symmetric tori. ``sweet spot''
$=(\theta/\pi,K,J_2)=(0.325,0.20,0.40)$, Eq.~(\ref{eq:sweetspot}); ``simple''
$=(0.5,\,0.50,\,0)$; ``$J_1$--$J_2$--$K$'' $=(0.5,\,0.25,\,0.40)$. All doublet
members are singlets. Sector coverage is complete for $N=16,20$; for $N=26$ the
eight $C_4$-invariant sectors were diagonalized in full and the lowest level of
each of the six $C_4$-stars of non-invariant momenta computed separately (the
lowest lies at $\Delta E=1.483$ and $1.168$ for the two $\theta=\pi/2$ points,
in both cases far above the doublet), so the quoted isolations are the true ones.}
\label{tab:clusters}
\end{table}

The results are collected in Table~\ref{tab:clusters}, and three conclusions
follow.

\paragraph{(i) The doublet quantum numbers are cluster-dependent.} For $N=16$ and
$20$ ($N/2$ even) both members sit at $\mathbf{k}=(0,0)$ with $C_4=\pm1$, the
familiar $A/B$ pair of the main text. For $N=26$ ($N/2$ odd) the pair splits
between $(\pi,\pi)\,A$ and $(0,0)\,E_1$. Searching for the ``$\mathbf{k}=0$,
$A/B$'' signature on the $26$-site cluster would therefore miss the doublet
entirely. Note in this context that the two $C_4=\pm i$ sectors are \emph{not}
degenerate: since $\HK$ breaks time reversal, no antiunitary operation relates
$\chi=+i$ to $\chi=-i$, so $E_1$ and $E_2$ are inequivalent one-dimensional
irreps and only one of them belongs to the doublet. At $N=26$, measured from the
ground state, $E_1$ lies at $0.0223$ (sweet spot), $0.0737$ (simple) and $0.0537$
($J_1$--$J_2$--$K$), while $E_2$ lies at $0.1532$, $1.1361$ and $0.9789$
respectively.

\paragraph{(ii) Both purely chiral points survive, and at $N=26$ they are the
better ones.} The isolation-to-splitting ratio at $N=26$ is $11.7$ for the simple
model and $17.1$ for $J_1$--$J_2$--$K$, against $5.9$ for the sweet spot on the same
cluster. In particular the absence of any convincing CSL signature at
$\theta=\pi/2$, $J_2=0$ on the $4\times4$ torus ($\Delta_{AB}=1.21$) --- the
observation that motivated the fine-tuning of Eq.~(\ref{eq:sweetspot}) in the first
place --- is a small-cluster artefact rather than a property of the model. This is
the result that matters for Appendix~\ref{app:pulses}: the real part of the cyclic
term is not needed, and $J_2$ is a refinement rather than a requirement.

\paragraph{(iii) The sweet spot is a $4\times4$-optimized point, and should be read
as such.} Its isolation ratio falls from $12.5$ at $N=16$ to $1.4$ at $N=20$ and
$5.9$ at $N=26$; at $N=16$ and $20$, in fact, \emph{all three} parameter sets look
unconvincing by one measure or another, so neither of these clusters resolves the
question of which point is best in the thermodynamic limit. ($N=20$ is a
particularly poor diagnostic here: at $J_2=0$, $\theta=\pi/2$ the $(0,0)A$ and
$(0,0)B$ levels simply cross near $K\simeq0.451$, and the apparent doublet is
strongly $K$-dependent away from that crossing --- $\Delta_{AB}=0.271$ at $K=0.48$
and $0.437$ at $K=0.50$ --- unlike the robust $26$-site doublet.)

We stress that none of this weakens the Floquet analysis of the main text, because
that analysis does not require the sweet spot to be optimal at every size. What it
requires is a cluster on which the static doublet is cleanly quasi-degenerate and
spectrally isolated, so that the fate of a \emph{known} CSL doublet under the drive
can be tracked unambiguously; Eq.~(\ref{eq:sweetspot}) supplies exactly that on the
$4\times4$ torus, which is also the largest cluster on which the full Floquet
propagator can be diagonalized in every symmetry sector at arbitrary $T$. Table
\ref{tab:clusters} does, however, identify the $\theta=\pi/2$ points as the natural
targets for both larger-scale follow-up studies and experimental realization, and it
is those that Appendix~\ref{app:pulses} takes as its reference.

\section{Full quasienergy spectrum and folding}
\label{app:fullspec}

\begin{figure}[t]
\centering
\includegraphics[width=0.8\textwidth]{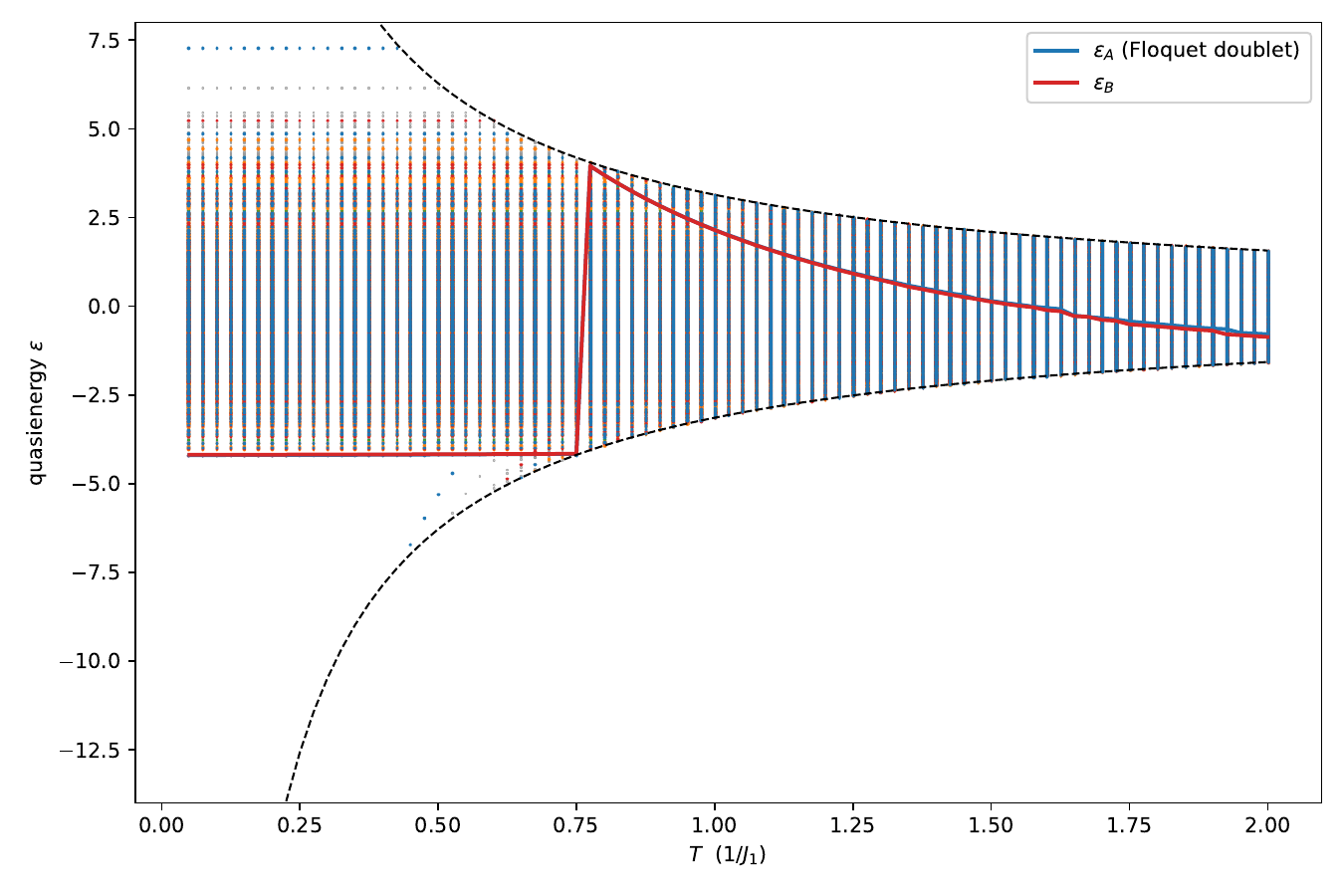}
\caption{Full Floquet quasienergy spectrum $\varepsilon_n$ (all $12870$
levels of the $S^z=0$ sector, dots) versus the drive period $T$, at the sweet
spot $(\theta/\pi,K,J_2)=(0.325,0.20,0.40)$. Dashed lines: the Floquet-zone
edges $\pm\pi/T$. Dots are coloured by symmetry sector: blue $A$ and red $B$ (the
two $\mathbf{k}=0$, $C_4=\pm1$ sectors that host the doublet), orange
$\mathbf{k}=(\pi,\pi)$, green the two $\mathbf{k}=0$, $C_4=\pm i$ sectors, grey all
remaining momenta. The doublet itself is the pair of solid lines, $\varepsilon_A$
(blue) and $\varepsilon_B$ (red). Figure~\ref{fig:doublet} (left) is obtained by
tracking the splitting and isolation of this doublet on the quasienergy circle.}
\label{fig:fullspec}
\end{figure}

For completeness, Fig.~\ref{fig:fullspec} shows the \emph{complete} quasienergy
spectrum of $\UF(T)$ at the sweet spot --- all $12870$ levels of the $S^z=0$
sector --- as a function of the drive period $T$, from which the doublet
tracking of Fig.~\ref{fig:doublet} (left) is extracted. Each level is folded
into the first Floquet zone $(-\pi/T,\pi/T]$, whose edges $\pm\pi/T$ are the
dashed hyperbolas; the topological doublet members $\varepsilon_A$ (blue) and
$\varepsilon_B$ (red) are highlighted. The figure makes the two folding scales
of Eqs.~(\ref{eq:wfold})--(\ref{eq:wres}) directly visible: at small $T$ the
whole spectrum sits inside the quasienergy zone and the doublet lies flat at the unfolded
bottom $\varepsilon_A\simeq\varepsilon_{\min}=-4.196$; near
$T_{\mathrm{fold}}=0.432$ the top of the band reaches the upper zone edge and
begins to re-enter from below; and at $T_{\mathrm{res}}=0.548$ the first folded
level reaches the doublet. Beyond $T\simeq0.77$ the (folded) $\varepsilon_B$
member wraps around the zone and reappears at the top, producing the apparent
jump of the red curve --- a purely kinematic effect of the modular
($\bmod\ 2\pi/T$) definition of the quasienergy, with no counterpart in the
average-energy ordering of Sec.~\ref{sec:ae}.

\section{Entanglement spectrum computed from the PEPS transfer operator}
\label{app:es}

This appendix provides some technical details on the calculation of the ES on an infinite half-cylinder carried out  in Sec.~\ref{sec:peps}.
The optimal site tensor is closed on an infinite cylinder of
circumference $N_v$; a single column of the double tensor
$\mathbb{E}=\sum_s {\mathcal A}^s\otimes \bar {\mathcal A}^s$ defines the transfer operator of dimension
$(D^2)^{N_v}$, which is block diagonal in the $\mathbb{Z}_2$ parity (even/odd
number of spin-$\tfrac12$ virtual states) of the boundary. Its leading right
eigenvectors $\sigma_R^{\mu}$ in each parity sector $\mu$, obtained by Arnoldi
iteration (the left ones follow from $\sigma_L=\sigma_R^{\mathsf T}$, which we
verified numerically to $10^{-6}$ against independently computed left fixed
points), yield the reduced density operator
$\rho^{(\mu)}=\sqrt{(\sigma_R^{\mu})^{\mathsf T}}\,\sigma_L^{\mu}
\sqrt{(\sigma_R^{\mu})^{\mathsf T}}$ acting on the virtual boundary space
$(\tfrac12\oplus 0)^{\otimes N_v}$ \cite{psa2016}. Each level of
$-\ln\rho^{(\mu)}$ is labeled by the edge momentum $k$ (lower case to avoid confusion with the
chiral coupling $K$) and by the total boundary spin $S$. Besides the parity
$p$, the sectors are refined by the $\mathbb{Z}_2$ flux $f$: inserting the gauge
string $g=\mathrm{diag}(-1,-1,+1)$ (doubled, $g\otimes\bar g$) on one ring bond
of the transfer operator selects $f=\pi$. In a flux sector the
transfer operator commutes not with the bare ring translation
${\mathcal T}$ but with the \emph{magnetic} translation
$\tilde{\mathcal T} = {\mathcal G}\,{\mathcal T}$, where the gauge operator
${\mathcal G}$ applies the signs $g\otimes\bar g$ on the leg crossing the
string (here ${\mathcal T}$ and ${\mathcal G}$ act on the boundary Hilbert
space); since
$\tilde{\mathcal T}^{N_v}$ equals the boundary $\mathbb{Z}_2$ parity, the momenta are
integer (half-integer) in units of $2\pi/N_v$ in the even (odd) parity sector.
Importantly, we find that the identification of the physical sectors depends
on the circumference modulo 4: at $N_v\equiv0\ (\mathrm{mod}\ 4)$ the vacuum
tower resides in $(p,f)=(\mathrm{even},0)$, whereas at
$N_v\equiv2\ (\mathrm{mod}\ 4)$ it resides in $(\mathrm{even},\pi)$ --- at
$N_v=6$ the $f=\pi$ even fixed point dominates the whole transfer spectrum
(leading eigenvalue $\lambda_{e,\pi}=7.30\times10^{-6}$ vs $\lambda_{e,0}=3.85\times10^{-6}$ at $T=0$),
while the subdominant $(\mathrm{even},0)$ spectrum carries the content of a
twisted sector, with a quasi-degenerate $(0)\oplus(1)$ base and an early
$S{=}2$ level that do not belong to the $SU(2)_1$ vacuum module. The
\emph{spinon} sector follows the complementary assignment: at $N_v=6$ it is
$(\mathrm{odd},0)$ --- with plain-translation, hence integer, momenta ---
whose spectrum displays the exact $SU(2)_1$ $j{=}\tfrac12$ tower content,
whereas the $(\mathrm{odd},\pi)$ spectrum (half-integer momenta) is
non-conformal. Both physical towers therefore carry integer momenta at
$N_v=6$; we checked this parity/flux pairing as well on the reference chiral
point 
 of
Refs.~\cite{poilblanc2015,psa2016}, where the physical sectors also dominate
their respective parity blocks and reproduce the towers with a wide
entanglement gap.

Finite-size (and finite-$D$) deviations from the ideal conformal towers are
also cleanly resolved. In the vacuum sector at $\Delta k=4$ the expected
$(2){+}2(1){+}2(0)$ content is only partially recovered: the $(0)$ and $(2)$
multiplets lie on the branch at $\xi\simeq6.3$, while the remaining members
are spread up to $\xi\simeq9$--$10$, where they merge with generic
(non-universal) boundary levels, visible, e.g., above $\xi\simeq9$ at
$\Delta k=0$. In the spinon sector, the finite entanglement gap of the
optimized tensors shows up as a non-universal extra $(\tfrac12)$ level at
$\Delta k=1$, at $\xi\simeq4.4$--$4.5$, above the tower states; at the ideal
reference point 
of
Refs.~\cite{poilblanc2015,psa2016} this intruder is pushed up to
$\xi\simeq5.9$, leaving the celebrated well-separated chiral branch.

\section{A digital pulse sequence for the drive}
\label{app:pulses}

The two-step drive (\ref{eq:UF}) was introduced above as a theoretical
construction, but it is in fact well matched to the gate sets of programmable
quantum simulators. This appendix gives an explicit realization of $\UF(T)$ built
\emph{entirely} from XY-type two-spin interactions on the four bond colours of the
square lattice, together with global single-qubit rotations --- the resources
natively available on Rydberg-atom, polar-molecule and NV platforms. That this
combination is already enough to Floquet-engineer generic spin-exchange models,
including anisotropic and Dzyaloshinskii--Moriya terms, has been demonstrated
for Rydberg arrays using only experimentally established capabilities
\cite{nishad2023}; the sequence below is an application of the same philosophy
to the specific target (\ref{eq:UF}). Throughout we
take the purely chiral point $\theta=\pi/2$ justified in
Appendix~\ref{app:larger}, so that
\begin{equation}
\HAF = J_1\!\!\sum_{\langle ij\rangle}\!\!\mathbf{S}_i\!\cdot\!\mathbf{S}_j
+ J_2\!\!\sum_{\langle\langle ij\rangle\rangle}\!\!\mathbf{S}_i\!\cdot\!\mathbf{S}_j ,
\qquad
\HK = K\sum_p i\big(P_p-P_p^{-1}\big)
= 2K\!\!\sum_{\text{corner triangles}}\!\!\!\!
\mathbf{S}_i\!\cdot\!(\mathbf{S}_j\times\mathbf{S}_k),
\end{equation}
i.e.\ $\HK$ is the uniform-flux scalar-chirality term. Label by $A$ the
checkerboard sublattice ($m+n$ even) and define the four bond colours
$a=(i,i+\hat e_x)$, $b=(i,i+\hat e_y)$, $c=(i,i-\hat e_x)$, $d=(i,i-\hat e_y)$
with $i\in A$; each is a perfect matching, i.e.\ a set of disconnected bonds.
Write $h_l=\sum_{\langle ij\rangle\in l}\mathbf{S}_i\!\cdot\!\mathbf{S}_j$,
$U_l(\phi)=e^{-i\phi h_l}$, $h_{\mathrm{NN}}=\sum_l h_l$, and
$h_x=h_a-h_c$, $h_y=h_b-h_d$.

\paragraph{The elementary gate is exact.} On a single bond let
$A=\sigma^x_i\sigma^x_j$, $B=\sigma^y_i\sigma^y_j$, $C=\sigma^z_i\sigma^z_j$.
Passing from $A$ to $B$ anticommutes on site $i$ \emph{and} on site $j$, so the two
sign changes cancel and $[A,B]=[B,C]=[C,A]=0$. Since
$h_{XY}=\tfrac14(A+B)$, $h_{YZ}=\tfrac14(B+C)$, $h_{XZ}=\tfrac14(A+C)$ and
$h_{XY}+h_{YZ}+h_{XZ}=2\,\mathbf{S}_i\!\cdot\!\mathbf{S}_j$, the three pulse
Hamiltonians commute exactly and
\begin{equation}
e^{-i\tau h_{XZ}}\,e^{-i\tau h_{YZ}}\,e^{-i\tau h_{XY}}
=\exp\!\big(\!-2i\tau\,\mathbf{S}_i\!\cdot\!\mathbf{S}_j\big)
\label{eq:gate}
\end{equation}
holds \emph{identically}, for arbitrary $\tau$ and independently of the ordering
(verified to $2\times10^{-16}$). The bonds within a colour being disjoint, three XY
pulses therefore give one exact Heisenberg layer $U_l(\phi)$, $\phi=2J\tau$, with no
Trotter error of any order: the entire Floquet structure below comes from the
\emph{ordering of the colours}, not from the decomposition. Two remarks. Since
$\mathbf{S}_i\!\cdot\!\mathbf{S}_j$ is $SU(2)$ invariant, the sign of a Heisenberg
gate cannot be flipped by single-qubit rotations --- but it need not be, as
$U_l(-\phi)=U_l(2\pi-\phi)$ up to a global phase, so a ``negative'' layer is just a
longer positive pulse and no sign-tunable coupling is required. And with unequal
durations the same argument gives, still exactly, an XYZ exchange with
$J_x=J(\tau_{XY}+\tau_{XZ})$ and cyclic permutations, the isotropic point
corresponding to equal durations.

\begin{figure}[t]
\centering
\includegraphics[width=0.95\textwidth]{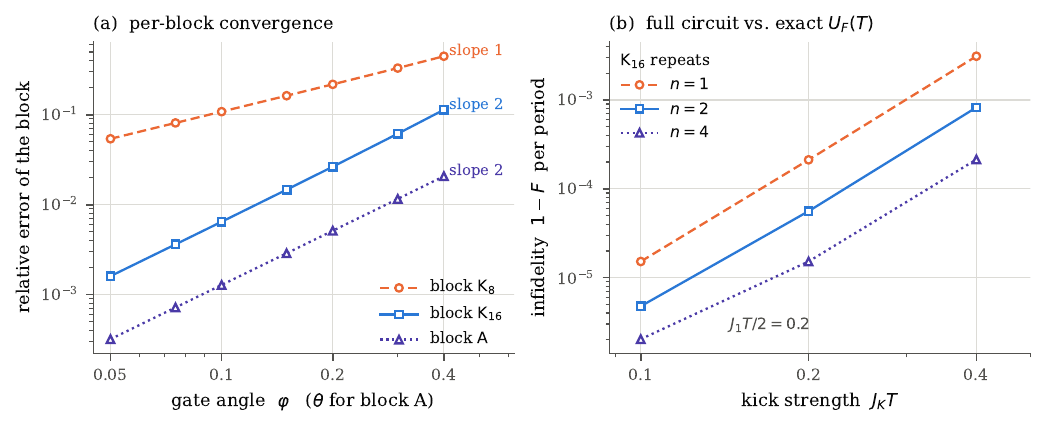}
\caption{(a) Relative error of each block against its target, on a $4\times2$ open
cluster ($8$ spins). The bare eight-layer chiral block $\mathrm{K}_8$ converges
linearly in the gate angle; the sign-reversed pair $\mathrm{K}_{16}$ of
Eq.~(\ref{eq:blockK}) gains a full order ($L_3=0$), matching the accuracy class of
the palindromic block $\mathcal{A}$ of Eq.~(\ref{eq:blockA}). (b) Infidelity per
period of the complete circuit $[\mathcal{K}(\varphi)]^n\mathcal{A}(\alpha)$
against the exact $\UF(T)$ of Eq.~(\ref{eq:UF}), at $J_1T/2=0.2$. Repeating the
chiral block $n$ times at fixed total kick improves the infidelity as $1/n^2$.}
\label{fig:conv}
\end{figure}

\paragraph{Block A: the Heisenberg half-step.} The four $h_l$ do not commute, so
$e^{-i\alpha h_{\mathrm{NN}}}$ with $\alpha=J_1T/2$ must be Trotterized. The naive
product $U_dU_cU_bU_a$ is the wrong choice: its leading Magnus correction is
\emph{pure scalar chirality}, exactly the quantity we wish to control
independently. The remedy is a time-symmetric product,
\begin{equation}
\mathcal{A}(\alpha)=U_a(\tfrac\alpha2)\,U_b(\tfrac\alpha2)\,U_c(\tfrac\alpha2)\,
U_d(\alpha)\,U_c(\tfrac\alpha2)\,U_b(\tfrac\alpha2)\,U_a(\tfrac\alpha2),
\qquad \alpha=\frac{J_1T}{2}.
\label{eq:blockA}
\end{equation}
Being palindromic, all \emph{even} Magnus orders vanish; in particular $L_2=0$
identically, so
$\log\mathcal{A}=-i(\alpha h_{\mathrm{NN}}+\alpha^3X_3+O(\alpha^5))$ and this block
generates \textbf{no chirality at all}. The residual $X_3$ is purely $SU(2)$
invariant and non-chiral; computed exactly on the $4\times4$ torus it consists of a
bond-modulated nearest-neighbour shift with coefficients
$\{-\tfrac14,-\tfrac18,0,+\tfrac18\}$ on the four colours (a relative
$\delta J_1/J_1\le\alpha^2/4$), a generated second/third-neighbour exchange
$\alpha^2/24$, and four-spin ring exchange on $280$ supports. Numerically the
relative error of $\mathcal{A}$ against $e^{-i\alpha h_{\mathrm{NN}}}$ is
$0.13\,\alpha^2$ with fitted exponent $3.00$ [Fig.~\ref{fig:conv}(a)]. Cost: $7$
Heisenberg layers, $21$ XY pulses.

If the frustrating $J_2$ is wanted, it cannot come from the Trotter error --- the
generated $\alpha^2/24$ would require $\alpha=2.7$ to reach $0.3$, and it is
accompanied by an equal third-neighbour coupling, so this route is a dead end. It
must come from the lattice: the diagonal bonds split into four further perfect
matchings $e,f=\{(i,i+\hat e_x+\hat e_y)\}$ and $g,h=\{(i,i+\hat e_x-\hat e_y)\}$
(each direction split by the parity of $m$), and extending the palindrome to all
eight colours gives $15$ layers, still with $L_2=0$ and hence still no chirality.
There is a useful accident here: for a dipolar $1/r^3$ interaction --- the natural
one on all three platforms mentioned above --- the diagonal coupling is fixed by
geometry at $J_2/J_1=2^{-3/2}=0.354$, within $13\%$ of the value $0.40$ of
Eq.~(\ref{eq:sweetspot}), so running all eight colours with \emph{identical} pulse
durations realizes $J_2/J_1=0.354$ with no calibration at all.

\paragraph{Block K: the chiral kick.} Chirality is a three-spin operator and can
only arise from a commutator of two-body terms, so any block built from Heisenberg
layers produces it at \emph{second} order in the gate angle. Start from the
eight-layer signed block
\begin{equation}
\mathcal{S}(\varphi):\qquad
a^{+}\ c^{-}\ b^{+}\ d^{-}\ \big|\ c^{+}\ a^{-}\ d^{+}\ b^{-},
\qquad l^{\pm}\equiv U_l(\pm\varphi).
\end{equation}
Two features are deliberate: the $\pm$ signs make the time average vanish, so
$L_1=0$ and $a,c$ enter only through $h_x=h_a-h_c$; and the reversal of the inner
colour order in the second half ($acbd|cadb$, not $acbd|acbd$) cancels $[h_a,h_c]$
and $[h_b,h_d]$, removing the collinear three-spin terms. The result is
$L_2=\varphi^2\,i[h_x,h_y]$ with \emph{zero} non-chiral weight --- but a large
$L_3$. That odd term is removed by concatenating the block with its sign-reversed
copy $\bar{\mathcal S}(\varphi)\equiv\mathcal S(-\varphi)$: since
$L_n(\bar{\mathcal S})=(-1)^nL_n(\mathcal S)$ and $L_1=0$,
\begin{equation}
\log\big(\bar{\mathcal S}\,\mathcal S\big)
=(L_2-L_3)+(L_2+L_3)+\tfrac12\big[L_2-L_3,L_2+L_3\big]+\dots
=2L_2+O(\varphi^4),
\end{equation}
the cross term $[L_2,L_3]$ being $O(\varphi^5)$. Exact rational arithmetic confirms
that the sixteen-layer block has $L_1=0$, $L_2=2\varphi^2\,i[h_x,h_y]$ with zero
non-chiral weight, and $L_3=0$ \emph{exactly}. Using
$i[h_x,h_y]=\tfrac12\sum_p i(P_p-P_p^{-1})$,
\begin{equation}
\mathcal{K}(\varphi)=\bar{\mathcal S}(\varphi)\,\mathcal S(\varphi)
=\exp\Big[-i\,\varphi^{2}\sum_p i\big(P_p-P_p^{-1}\big)\Big]+O(\varphi^4),
\qquad \varphi=\sqrt{\tfrac{KT}{2}},
\label{eq:blockK}
\end{equation}
and reversing the block reverses the sign of $K$, hence the sense of the emergent
flux. Note the square root: the kick angle is intrinsically \emph{second} order in
a two-body gate angle. The relative error of $\mathcal{K}$ is
$0.65\,\varphi^2=0.33\,KT$ with fitted exponent $4.00$ [Fig.~\ref{fig:conv}(a)],
against $O(\varphi^3)$ for the bare eight-layer block. Just as important, the
residual barely leaks into a Heisenberg term: projected onto $h_{\mathrm{NN}}$ it is
$7\times10^{-5}$ relative at $\varphi=0.1$ and $5.5\times10^{-4}$ at $\varphi=0.2$,
so the kick does not renormalize $J_1$ and the two couplings of
Eq.~(\ref{eq:UF}) remain independent. Repeating the block $n$ times with
$\varphi=\sqrt{KT/2n}$ leaves the total kick unchanged and reduces the relative
error to $0.33\,KT/n$: one should increase $n$, never $\varphi$.

\paragraph{Assembly, cost and accuracy.} One Floquet period is
\begin{equation}
\UF(T)\;\simeq\;\big[\mathcal{K}(\varphi)\big]^{n}\,\mathcal{A}(\alpha),
\qquad \alpha=\frac{J_1T}{2},\qquad \varphi=\sqrt{\frac{KT}{2n}},
\end{equation}
costing $7+16n$ Heisenberg layers, i.e.\ $21+48n$ XY pulses ($15+16n$ layers and
$45+48n$ pulses if $J_2$ is included); the trailing $U_a(\alpha/2)$ of block A
merges with the leading $U_a(+\varphi)$ of block K, saving one layer per period.
Table~\ref{tab:e2e} gives the end-to-end comparison against the exact
Eq.~(\ref{eq:UF}) on an $8$-spin cluster, with
$F=|\mathrm{Tr}\,U^\dagger_{\mathrm{dig}}\UF|/\dim$. At the natural working point
$J_1T/2=0.2$, $KT=0.1$ a single chiral block already reproduces the exact
propagator to $1.5\times10^{-5}$ per period.

\begin{table}[h]
\centering
\begin{tabular}{cccc c}
\toprule
$J_1T/2$ & $KT$ & $n$ & $\varphi$ & $1-F$ \\
\midrule
$0.2$ & $0.1$ & $1$ & $0.224$ & $1.5\times10^{-5}$\\
$0.2$ & $0.1$ & $4$ & $0.112$ & $2.0\times10^{-6}$\\
$0.2$ & $0.2$ & $1$ & $0.316$ & $2.1\times10^{-4}$\\
$0.2$ & $0.2$ & $4$ & $0.158$ & $1.5\times10^{-5}$\\
$0.2$ & $0.4$ & $1$ & $0.447$ & $3.1\times10^{-3}$\\
$0.2$ & $0.4$ & $4$ & $0.224$ & $2.1\times10^{-4}$\\
\bottomrule
\end{tabular}
\caption{Infidelity per period of the digital circuit against the exact kicked
evolution (\ref{eq:UF}), on a $4\times2$ open cluster.}
\label{tab:e2e}
\end{table}

The asymmetry between the two blocks is intrinsic and worth stating plainly: the
Heisenberg half-step is \emph{first} order in the gate angle, hence cheap and
accurate, whereas the chiral kick is \emph{second} order, hence expensive and the
accuracy bottleneck. This is not an artefact of the particular sequence --- a
circuit of two-body gates reaches a three-spin operator only through a commutator.
It is also what makes the $J_1$--$J_2$--$K$ point of Appendix~\ref{app:larger}
attractive despite its extra layers: halving $K$ from $0.5$ to $0.25$ --- which the
frustrating $J_2$ permits, following the trend established in
Refs.~\cite{nielsen2013,poilblanc2017} --- halves the kick error, for a fixed cost
of $+24$ XY pulses in block A, whereas the $J_2=0$ point would need $n=2$ to reach
the same fidelity. Since the dipolar $J_2/J_1=0.354$ comes for free on the relevant
platforms, this is the cheaper way to buy accuracy. Note, in this respect, that
every operator generated at any order is an $SU(2)$ scalar --- the circuit commutes
with $\mathbf{S}_{\mathrm{tot}}$ exactly, by Eq.~(\ref{eq:gate}) --- so no
Dzyaloshinskii--Moriya term and no exchange anisotropy is produced, at any order and
for any angles.

Two caveats close this appendix. First, the dominant
systematic at the parameters of Table~\ref{tab:e2e} is the $\alpha^3$ bond
modulation of $J_1$ from block A ($\le\alpha^2/4$, i.e.\ $1\%$ at $\alpha=0.2$)
together with the generated $J_2/J_1=\alpha^2/24$; both break $C_4$ mildly and
favour valence-bond order, and both are removable by a higher-order Suzuki product
at the cost of more layers, or absorbable by redefining the target couplings.
Second, Eq.~(\ref{eq:gate}) assumes the bonds of the \emph{other} colours are
genuinely switched off during each sub-pulse; residual always-on couplings, not the
decomposition, are what will limit the fidelity in practice, and quantifying them
requires a platform-specific model. Establishing that the topological signatures of
the main text --- rather than the gate fidelity alone --- survive the digital
approximation on a $4\times4$ torus is left for future work.

\end{document}